\documentstyle[12pt]{article}
\input psfig.sty
\input epsfig.sty
\def\nn{\noindent}
\def\Re{{\cal R \mskip-4mu \lower.1ex \hbox{\it e}\,}}
\def\Im{{\cal I \mskip-5mu \lower.1ex \hbox{\it m}\,}}
\def\ie{{\it i.e.}}
\def\eg{{\it e.g.}}

\def\ibid{{\it ibid}.}
\def\mpl{\ifmmode \overline M_{Pl}\else $\overline M_{Pl}$\fi}
\def\sub#1{_{\lower.25ex\hbox{$\scriptstyle#1$}}}
\def\tev{\,{\ifmmode\mathrm {TeV}\else TeV\fi}}
\def\gev{\,{\ifmmode\mathrm {GeV}\else GeV\fi}}
\def\mev{\,{\ifmmode\mathrm {MeV}\else MeV\fi}}
\def\to{\rightarrow}

\def\subw{_{\rm w}}
\def\mh{\ifmmode m\sbl H \else $m\sbl H$\fi}
\def\mch{\ifmmode m_{H^\pm} \else $m_{H^\pm}$\fi}
\def\mt{\ifmmode m_t\else $m_t$\fi}
\def\mc{\ifmmode m_c\else $m_c$\fi}
\def\mz{\ifmmode M_Z\else $M_Z$\fi}
\def\mw{\ifmmode M_W\else $M_W$\fi}
\def\mws{\ifmmode M_W^2 \else $M_W^2$\fi}
\def\mhs{\ifmmode m_H^2 \else $m_H^2$\fi}   
\def\mzs{\ifmmode M_Z^2 \else $M_Z^2$\fi}
\def\mts{\ifmmode m_t^2 \else $m_t^2$\fi}
\def\mcs{\ifmmode m_c^2 \else $m_c^2$\fi}
\def\mchs{\ifmmode m_{H^\pm}^2 \else $m_{H^\pm}^2$\fi}
\def\ztwo{\ifmmode Z_2\else $Z_2$\fi}
\def\zone{\ifmmode Z_1\else $Z_1$\fi}
\def\mtwo{\ifmmode M_2\else $M_2$\fi}
\def\mone{\ifmmode M_1\else $M_1$\fi}
\def\tb{\ifmmode \tan\beta \else $\tan\beta$\fi}
\def\xw{\ifmmode x\subw\else $x\subw$\fi}
\def\ch{\ifmmode H^\pm \else $H^\pm$\fi}
\def\lum{\ifmmode {\cal L}\else ${\cal L}$\fi}
\def\inpb{\,{\ifmmode {\mathrm {pb}}^{-1}\else ${\mathrm {pb}}^{-1}$\fi}}
\def\infb{\,{\ifmmode {\mathrm {fb}}^{-1}\else ${\mathrm {fb}}^{-1}$\fi}}
\def\epem{\ifmmode e^+e^-\else $e^+e^-$\fi}
\def\ppb{\ifmmode \bar pp\else $\bar pp$\fi}
\def\bsg{\ifmmode B\to X_s\gamma\else $B\to X_s\gamma$\fi}
\def\bsll{\ifmmode B\to X_s\ell^+\ell^-\else $B\to X_s\ell^+\ell^-$\fi}
\def\bstt{\ifmmode B\to X_s\tau^+\tau^-\else $B\to X_s\tau^+\tau^-$\fi}
\def\lamt{\ifmmode \tilde\lambda\else $\tilde\lambda$\fi}
\def\shat{\ifmmode \hat s\else $\hat s$\fi}
\def\that{\ifmmode \hat t\else $\hat t$\fi}
\def\uhat{\ifmmode \hat u\else $\hat u$\fi}

\newskip\zatskip \zatskip=0pt plus0pt minus0pt
\def\matth{\mathsurround=0pt}

\def\atversim#1#2{\lower0.7ex\vbox{\baselineskip\zatskip\lineskip\zatskip
  \lineskiplimit 0pt\ialign{$\matth#1\hfil##\hfil$\crcr#2\crcr\sim\crcr}}}

\renewcommand{\thefootnote}{\fnsymbol{footnote}}

\begin{document} \begin{titlepage} 
\rightline{\vbox{\halign{&#\hfil\cr
&SLAC-PUB-8635\cr
&FERMILAB-PUB-00/286-T\cr
&October 2000\cr}}}
\begin{center}

{\Large\bf Signals for Non-Commutative Interactions at \\
Linear Colliders}
\footnote{Work supported by the Department of 
Energy, Contract DE-AC03-76SF00515}
\medskip

\normalsize 
{\bf  JoAnne L. Hewett$^{a,b}$, Frank J. Petriello$^a$, and Thomas G. 
Rizzo$^a$}
\vskip .3cm
$^a$Stanford Linear Accelerator Center \\
Stanford University \\
Stanford CA 94309, USA\\
\vskip .1cm
$^b$Fermi National Accelerator Laboratory\\
Batavia, IL 60510, USA\\
\vskip .3cm
\end{center}

\begin{abstract} 
Recent theoretical results have demonstrated that
non-commutative geometries naturally appear within the context of
string/M-theory.  One consequence of this possibility is that
QED takes on a non-abelian nature due to the introduction of
3- and 4-point functions.  In addition, each QED vertex acquires
a momentum dependent phase factor.  We parameterize the effects of 
non-commutative space-time co-ordinates and show that they lead to
observable signatures in several $2\to 2$ QED processes in \epem\ 
collisions.  In particular, we examine pair annihilation, Moller 
and Bhabha scattering, as well as $\gamma\gamma\to\gamma\gamma$ 
scattering and show that non-commutative scales of order a TeV can 
be probed at high energy linear colliders.
\end{abstract} 

\renewcommand{\thefootnote}{\arabic{footnote}} \end{titlepage}


\section{Introduction and Background}

Although the full details of string/M-theory have yet to be unraveled,
this theoretical effort has  inspired a number of ideas over the years 
which have had significant impact on the phenomenology of particle physics. 
Two such examples are given by the string-inspired 
$E_6$ models of the late 1980's{\cite {physrep}} and the ongoing endeavor in 
building realistic and testable models from theories which have additional 
space-time dimensions{\cite {big}}. Most recently, a resurgence 
of interest in non-commutative quantum field theory (NCQFT) and its 
applications{\cite {basic}} has developed within the context of string theory.
Of course non-commutative theories are also interesting in their own right.
However, it has yet to be explored whether they  have any connection with 
the physics 
of the Standard Model (SM) or whether their effects could be observable in 
laboratory experiments. It is the purpose of this paper to begin to address 
these questions. 

An exhaustive introduction to NCQFT is beyond the scope of the present 
treatment, hence we will simply outline some of the basics of the theory 
as well as some results which are relevant to the phenomenological analysis 
that follows.  We will see that NCQFT results in modifications to QED which
can be probed in $2\to 2$ processes in \epem\ collisions.

The essential idea of NCQFT is a generalization of the usual d-dimensional 
space, $R^d$, associated with commuting space-time coordinates to one which 
is non-commuting, $R^d_\theta$. In such a space the conventional 
coordinates are represented by operators which no longer commute, \ie, 
\begin{equation}
[\hat X_\mu,\hat X_\nu]=i\theta_{\mu\nu}\equiv
{i\over {\Lambda_{NC}^2}}c_{\mu\nu}\,.
\end{equation}
In the last equality we have parameterized the effect in terms of an overall 
scale $\Lambda_{NC}$, which characterizes the threshold where non-commutative 
(NC) effects become relevant, and a real antisymmetric matrix $c_{\mu\nu}$, 
whose  dimensionless elements are presumably of order unity. 
From our point of view the role
of the NC scale $\Lambda_{NC}$ can be compared to that of
$\hbar$ in conventional Quantum Mechanics which represents the level of 
non-commutativity between coordinates and momenta.
{\it A priori} the scale $\Lambda_{NC}$ can take any value, perhaps the 
most likely being of order the Planck scale \mpl.  However, given the
possibility\cite{big} of the onset of stringy effects at the TeV scale, and
that values of the scale where gravity becomes strong
in models with large extra dimensions can be of order a TeV, it is
feasible that NC effects could also set in at a TeV.  Here, we adopt this 
spirit and consider the possibility that $\Lambda_{NC}$ may not lie far above
the TeV scale.

Note that the matrix $c_{\mu\nu}$ 
is {\it not} a tensor since its elements are identical in all reference 
frames. This leads immediately to a violation of Lorentz invariance which is 
quite different than that discussed most often in the 
literature\cite{violate} since 
it sets in only at energies of order $\Lambda_{NC}$. As we 
will see below, this violation will take the form of dimension-8 operators for 
the processes  we consider and is thus highly suppressed at low energies. 
In addition, there exists a more than superficial relation between the
anti-symmetric matrix $c_{\mu\nu}$ and the Maxwell field strength tensor
$F_{\mu\nu}$ as  NCQFT arises in string theory\cite{embed} through 
the quantization of
strings, described by the low energy excitations of D-branes 
in the presence of background electromagnetic fields.  The space-time
components, $c_{0i}$, thus define the direction of a background {\bf E} field,
while the space-space components, $c_{ij}$, describe the direction of a
background magnetic or string {\bf B} field.  Geometrically, 
we can then think of $c_{0i}$ and $c_{ij}$ as two 3-vectors that point in a 
specific pair of preferred directions in the laboratory frame.  Theories with 
$c_{0i}(c_{ij})\neq 0$ are usually referred to as space-time(space-space) 
non-commutative. 

NCQFT can be phrased in terms of conventional commuting QFT through the 
application of the Weyl-Moyal correspondence{\cite {follow}}
\begin{equation}
\hat A(\hat X) \longleftrightarrow A(x)\,,   
\end{equation}
where $A$ represents a quantum field with $\hat X$ being the set of  
non-commuting coordinates and $x$ corresponding to the commuting set.  However,
in formulating  NCQFT, one must be careful to preserve orderings
in expressions such as $\hat A(\hat X)\hat B(\hat X)$.  This is accomplished
with the introduction of a star product, $\hat A(\hat X)\hat B(\hat X) =
A(x)*B(x)$, where the effect of the commutation relation is absorbed into
the star.  Making the Fourier transform pair
\begin{eqnarray}
\hat A(\hat X) &=& {1\over {(2\pi)^{d/2}}}\int~d\alpha e^{i\alpha \hat X}
~a(\alpha)\nonumber \\
a(\alpha) &=& {1\over {(2\pi)^{d/2}}}\int~dx e^{-i\alpha x}~A(x)\,,
\end{eqnarray}
with $x$ and $\alpha$ being real n-dimensional variables, allows us to write 
the product of two fields as 
\begin{eqnarray}
\hat A(\hat X) \hat B(\hat X) &=& {1\over {(2\pi)^{d}}}\int~d\alpha d\beta 
e^{i\alpha \hat X}~a(\alpha)e^{i\beta \hat X}~b(\beta)\nonumber \\
&=& {1\over {(2\pi)^{d}}}\int~d\alpha d\beta ~e^{i(\alpha+\beta)\hat X-
{1\over {2}}\alpha^\mu \beta^\nu[\hat X_\mu,\hat X_\nu]}a(\alpha)b(\beta)\,.
\end{eqnarray}
We thus have the correspondence 
\begin{equation}
\hat A(\hat X) \hat B(\hat X) \longleftrightarrow A(x)*B(x)\,,   
\end{equation}
provided we identify
\begin{equation}
A(x)*B(x)\equiv \bigg[e^{{i\over {2}}\theta_{\mu\nu}\partial_{\zeta\mu}
\partial_{\eta\nu}}A(x+\zeta)B(z+\eta)\bigg]_{\zeta=\eta=0}\,. 
\end{equation}
Note that to leading order in $\theta$ the $*$ product is given by
\begin{equation}
A(x)*B(x)=AB+{i\over {2}}\theta^{\mu\nu}\partial_\mu A\partial_\nu B
+{\cal O}(\theta^2)\,. 
\end{equation}
Hence the non-commutative version of an action for 
a quantum field theory 
can be obtained from the ordinary one by replacing the products of fields 
by star products. In doing so it is useful to define a generalized commutator, 
known as the Moyal bracket, for two quantities $S,T$ as
\begin{equation}
[S,T]_{MB}=S*T-T*S\,,
\end{equation}
so that the Moyal bracket of any quantity with itself vanishes.
Note that the integration of a Moyal bracket of two quantities over all 
space-time vanishes, \ie, 
\begin{equation}
\int ~d^4x~[S(x),T(x)]_{MB}=0\,,
\end{equation}
which means they commute inside the integral. This can be generalized 
to show that the integral of a $*$ product of an arbitrary number of 
quantities is invariant under cyclic permutations in a manner similar to the 
trace of ordinary matrices. We also note that the Moyal bracket of two 
coordinates
\begin{equation}
[x_\mu,x_\nu]_{MB}=x_\mu*x_\nu-x_\nu*x_\mu\,,   
\end{equation}
mimics the operator commutation relation in Eq.~(1). 

Once the products of fields 
are replaced by $*$ products, infinite numbers of derivatives 
of fields can now appear in an action, 
implying that all such theories are non-local. This is not surprising since, 
in analogy with ordinary Quantum Mechanics, one now has a spacetime 
uncertainty relation
\begin{equation}
\Delta \hat X^\mu \Delta \hat X^\nu \geq {1\over {2}}|\theta^{\mu\nu}|\,. 
\end{equation}
Theories with $c_{0i}\neq 0$ have an additional problematic feature in 
that they generally do not 
have a unitary $S$-matrix{\cite {unitary}}, at least in perturbation theory, 
since an infinite number of time derivatives are involved in $*$ products. 
However, it has recently been shown\cite{newst} that it may be possible to 
unitarize the space-time case by combining the spacial NC Super Yang Mills
limit with a Lorentz transformation with finite boost velocity.
On the other hand, theories with only space-space non-commutativity, 
$c_{ij} \neq 0$, are unitary. 

\begin{figure}[htbp]
\centerline{
\psfig{figure=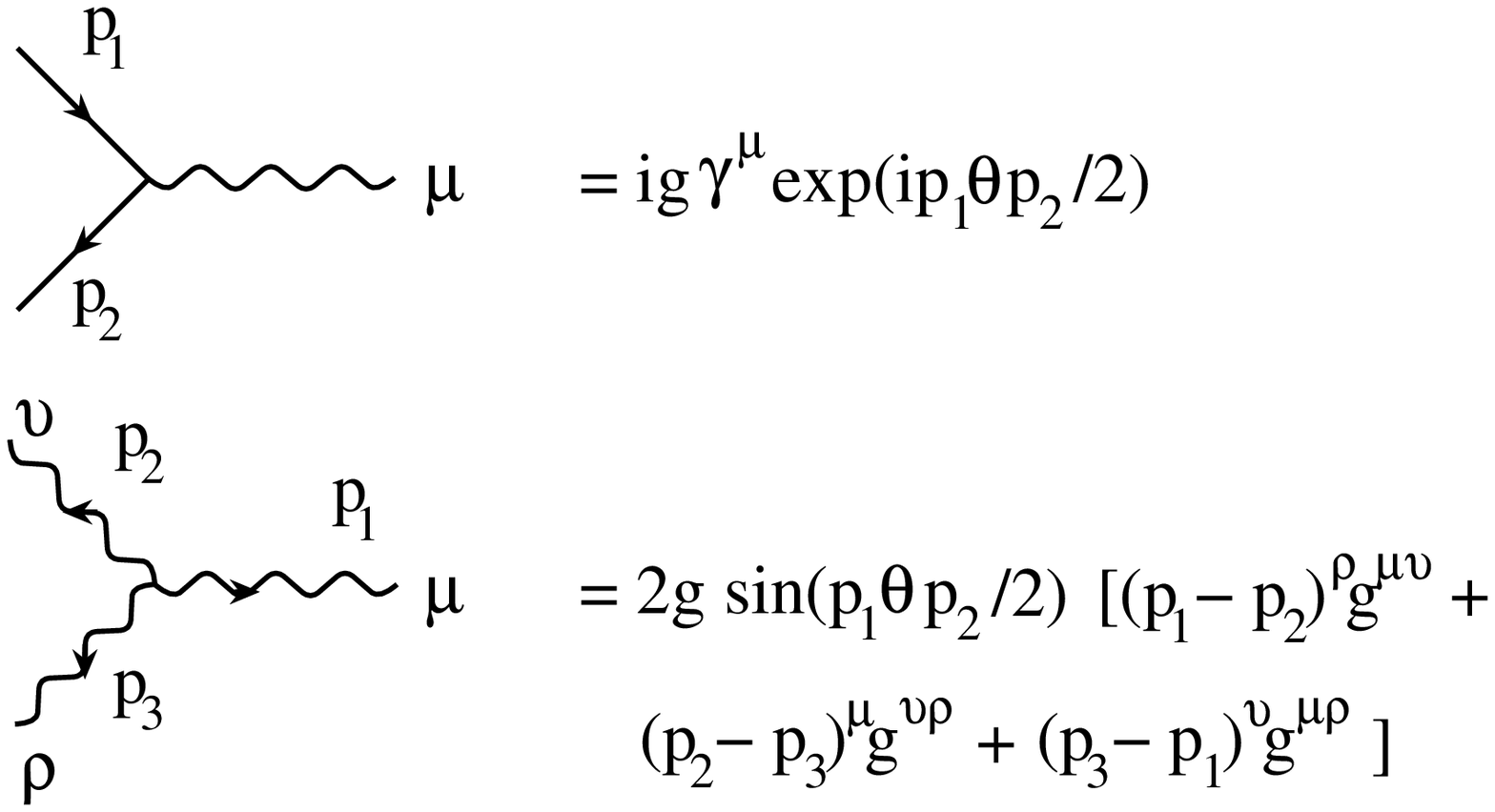,height=8.4cm,width=12.8cm,angle=0}}
\vspace*{-2.3cm}
\centerline{
\psfig{figure=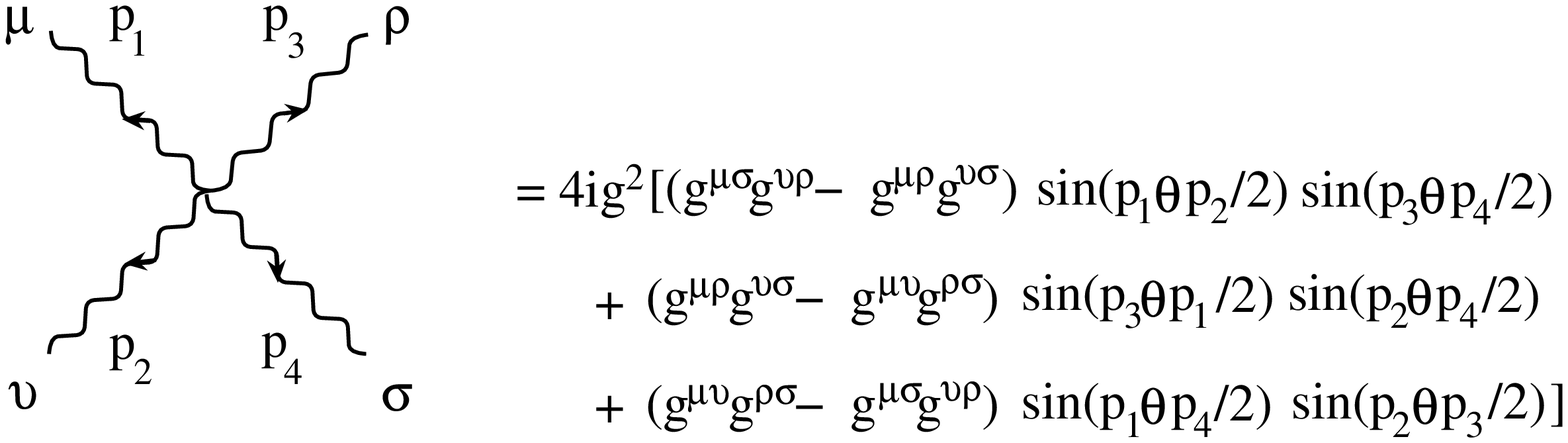,height=9.66cm,width=14.72cm,angle=0}}
\vspace*{-4.5cm}
\caption{Feynman rules of NCQED.}
\label{frules}
\end{figure}

There are a number of important results in NCQFT 
which we now state without proof, 
referring the interested reader to the original papers for 
detailed explanations. ($i$) Matsubara{\cite {matsu}} 
has shown that only the $U(n)$ matrix Lie algebra is closed under 
the Moyal bracket, thus non-commutative gauge theories can be constructed
only if they are based upon these gauge groups.  Hence in order to embed the
full SM in NCQFT, the usual Standard Model $SU(n)$ group factors must be 
extended to $U(n)$ groups.  However, due to the effective non-abelian nature 
of the Moyal 
brackets, these $U(n)$ groups cannot be simply decomposed into products of 
$SU(n)$ and $U(1)$ factors.
($ii$) There are indications that conventional renormalizable, gauge invariant 
field theories remain renormalizable and gauge invariant when generalized to 
non-commutative spacetimes{\cite {renorm}}, although a proof does not yet 
exist for theories which are spontaneously broken{\cite {bruce}}.
($iii$) Non-commutative QED, based on the group $U(1)$, has been studied 
by several groups{\cite {haya,ncqed}}; 
due to the presence of $*$ products and Moyal brackets, the theory takes 
on a non-abelian nature in that both 3-point and 4-point photon 
couplings are generated. The photonic part of the action is now
\begin{equation}
S_{NCQED}={-1\over {4}}\int~d^4x~F_{\mu\nu}*F^{\mu\nu}=
{-1\over {4}}\int~d^4x~F_{\mu\nu}F^{\mu\nu}\,,
\end{equation}
where the second equality follows from the commutativity of Moyal brackets 
under integration, shown in Eq.~(9). This action is gauge invariant 
under a local transformation $U(x)$ with $F_{\mu\nu}$ defined as
\begin{equation}
F_{\mu\nu}=\partial_\mu A_\nu-\partial_\nu A_\mu-i[A_\mu,A_\nu]_{MB}\,.
\end{equation}
The origin of the 3- and 4-point functions is now readily transparent.
Note that the photon's 2-point function is identical in commutative and NC
spaces because quadratic forms remain unchanged.  When performing the 
Fourier transformation of these new interactions into momentum space, the 
vertices pick up additional phase factors which are dependent upon the momenta 
flowing through the vertices.  We will see below that these kinematic phases 
will play an important role in the collider tests of NCQFT. 
($iv$) When fermions are added to the theory, covariant 
derivatives can only be constructed for fields of charge $0,\pm 1$. 
The structure of those 
derivatives for the $Q=\pm 1$ case is similar to that for fundamental and 
anti-fundamental representations in non-abelian theories. The covariant 
derivatives for neutral fields are either trivial (as in the case of  
abelian commutative $U(1)$ theory) or correspond to what would ordinarily be 
called the adjoint representation in the case of a commutative non-abelian 
theory. As before, the three-point function picks up an additional kinematic 
phase from the Fourier transformation of the interaction term into
momentum space.  This
is shown explicitly in Fig. \ref{frules}. The general form of the Feynman 
rules for NCQED can be found in Ref.{\cite {rules}}; the ones of relevance
to the processes considered in this paper are displayed in Fig. \ref{frules}.
($v$) NCQED with fermions and space-space non-commutativity has been 
shown{\cite {cp}} to be CP violating yet CPT conserving.

\noindent

Having now presented the basic formalism of NCQFT and subsequent modifications
to QED, in the following sections we examine the effects in several 
$2\to 2$ processes in \epem\
collisions, including pair annihilation, Moller and Bhabha scattering, as
well as in $\gamma\gamma\to\gamma\gamma$ scattering.
We will see that the lowest order correction
to the SM results for these transitions is given by dimension-8 operators.  
In addition, we find that an oscillatory azimuthal dependence
is induced in these processes due to the preferred direction in the
laboratory frame defined by the NC matrix $c_{\mu\nu}$.  In summary,
we will see that high energy linear colliders can probe non-commutative
scales of order a TeV.   

Before discussing our analysis for the specific processes considered here,
a few additional comments are in order regarding 
the observation of these non-commutative effects. First, as discussed above, 
the vectors $c^E_i=[c_{0i}]$ and $c^B_i=[\epsilon_{ijk}c_{jk}]$ point in fixed 
specific directions which are the same in all reference frames. 
In our analysis below we define the $z$-axis as that corresponding to the 
direction of the incoming particles in a fixed laboratory frame with the 
vectors {\bf c} having arbitrary 
components in that frame. Now, imagine a second laboratory at a different 
point on the surface of the Earth performing the same experiment. Clearly the 
co-ordinate systems of the two laboratories will be different, \ie, the 
beam directions and hence the $z$-axis will not be the same in the two 
locations. This implies that the experimentally determined values for the 
components of the {\bf c} vectors  
will differ at the two laboratory sites due to their locally chosen set of  
co-ordinates. Hence {\it both} laboratories must convert their local 
co-ordinates to a common frame, \eg, with respect to the rest frame of the 
3 degree K blackbody radiation or some other slowly varying astronomical
co-ordinate system, so that they would measure equivalent 
directions and magnitudes for {\bf c}. 
This translation of co-ordinates to a common frame will be 
necessary if we are to compare the results of multiple experiments for signals 
of non-commutativity.

In addition, even for a single experiment, the apparent directions of the 
{\bf c} vectors will vary with time due to the rotation of the Earth 
and its revolution about the Sun. While the actual {\bf c} vectors will 
always point to the 
same position on the sky, the co-ordinates of this position will vary 
continuously in the laboratory frame due to the Earth's motion.  (The
effects of galactic motion should be small during the life-span of any
given experiment.) Collider experiments will thus have to make use of 
astronometric techniques to continuously 
translate their laboratory co-ordinates to astronomical ones 
such that when  
events are recorded the relative orientation of the two frames would be 
accounted for. This should be a rather straightforward procedure 
for any future collider 
experiment to implement given that many non-accelerator based experiments 
already make use of these ideas.

Taking the Earth's motion into account is particularly important
for experiments which measure observable quantities that are odd in
{\bf c}, including for example, the $g-2$ of the muon\cite{haya}, 
the Lamb shift\cite{finns}, as well as other processes which are
linear\cite{pop} in the NC parameter.
If only the laboratory co-ordinates were employed, at least some of the
components of {\bf c} would average to zero over a sidereal day.  In the
cases we discuss below, the observables are even functions of {\bf c},
and while we would not obtain a null effect, time averaging would
result in a diminished sensitivity to $\Lambda_{NC}$.

\section{Moller Scattering}

For all of the scattering processes considered in this paper, except for
$\gamma\gamma\to\gamma\gamma$, we can define 
the momenta of the incoming, represented by $p_{1,2}$, and outgoing,
corresponding to $k_{1,2}$, particles in terms of the coordinates fixed in 
the laboratory as
\begin{equation}
\begin{array}{rcl}
p_1^\mu &=& {{\sqrt s}\over {2}}(1,-1,0,0)\\
k_1^\mu &=& {{\sqrt s}\over {2}}(1,-c_\theta,-s_\theta c_\phi,-s_\theta s_\phi)
\end{array}\qquad
\begin{array}{rcl}
p_2^\mu &=& {{\sqrt s}\over {2}}(1,1,0,0)\\
k_2^\mu &=& {{\sqrt s}\over {2}}(1,c_\theta,s_\theta c_\phi,s_\theta s_\phi)\,.
\end{array}
\label{momdep}
\end{equation}
Note that the ordering of the co-ordinates used in these definitions is 
given by $(t,z,x,y)$, so that the $z$-axis is along the beam 
direction as usual.  Using these definitions,
the bilinear products of these momenta with the matrix $c_{\mu\nu}$, which
appear in the Feynman rules of Fig. \ref{frules},
can be calculated to be 
\begin{eqnarray}
p_1\cdot c \cdot p_2 &=& {s\over {2}}c_{01}\nonumber \\
k_1\cdot c \cdot k_2 &=& {s\over {2}}[c_{01}c_\theta+c_{02}s_\theta c_\phi+
c_{03}s_\theta s_\phi]\nonumber \\
p_1\cdot c \cdot k_1 &=& {s\over {4}}[c_{01}(1-c_\theta)+(c_{12}-c_{02})
s_\theta c_\phi-(c_{03}+c_{31})s_\theta s_\phi]\nonumber \\
p_1\cdot c \cdot k_2 &=& {s\over {4}}[c_{01}(1+c_\theta)-(c_{12}-c_{02})
s_\theta c_\phi+(c_{03}+c_{31})s_\theta s_\phi]\nonumber \\
p_2\cdot c \cdot k_1 &=& {s\over {4}}[-c_{01}(1+c_\theta)-(c_{12}+c_{02})
s_\theta c_\phi-(c_{03}-c_{31})s_\theta s_\phi]\\
p_2\cdot c \cdot k_2 &=& {s\over {4}}[-c_{01}(1-c_\theta)+(c_{12}+c_{02})
s_\theta c_\phi+(c_{03}-c_{31})s_\theta s_\phi]\nonumber \,.
\label{dotprods}
\end{eqnarray}
We remind the reader of the fact that $a\cdot c \cdot a=0$ for 
all vectors $a$ due to the antisymmetry of the matrix $c$. 
Note that $c_{23}$ does not appear in any of the above expressions since we 
have defined the $z$-axis to be along the direction of the initial beams and 
there is no {\bf B} field associated non-commutative asymmetry relative to 
this direction. 

\begin{figure}[htbp]
\centerline{
\psfig{figure=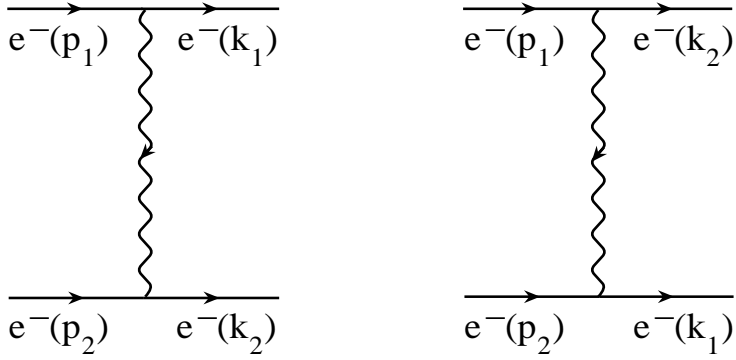,height=6cm,width=12cm,angle=0}}
\vspace*{-0.5cm}
\caption[*]{Feynman graphs contributing to Moller scattering with the 
exchanged particle corresponding to a photon and $Z$-boson.} 
\label{moller}
\end{figure}

The Feynman diagrams which mediate Moller scattering are displayed in
Fig. \ref{moller}.  In this case, the NC modifications correspond to the
kinematic phase which appears in each vertex.  The question here is how
to treat the $Z$-boson exchange contribution.  While 
NCQED is a well defined theory, it is not immediately clear how to extend it 
to the full SM in a naive way even if we are only interested in tree-level 
fermionic interactions.  Without such guidelines we see that there exist 
three possibilities: (i) the simplest case is if the $Z$ and 
photon have the same vertex structure as shown in Fig. \ref{frules}, (ii) the 
full theory and appropriate $Zf\bar f$ 
kinematic phase are more complex, or (iii) only $\gamma f\bar f$ 
vertices pick up kinematic phases. Clearly, as far as signatures of 
non-commutativity are concerned, cases one and three will be qualitatively 
similar.   Hence, for simplicity, we assume that the first possibility is 
realized. 

\vspace*{-0.5cm}
\nn
\vspace*{0.4mm}

Following the Feynman rules of Fig. \ref{frules} and the momentum labeling 
given in Fig. \ref{moller} we see that the $t$- and $u$-channel exchange 
graphs now pick up kinematic phases given by
\begin{eqnarray}
\phi_t &=& {1\over {2}}[p_1\cdot \theta \cdot k_1+p_2\cdot \theta \cdot 
k_2]\nonumber \\
\phi_u &=& {1\over {2}}[p_1\cdot \theta \cdot k_2+p_2\cdot \theta \cdot k_1]\,.
\end{eqnarray}
Clearly, only the interference terms between the 
$t$- and $u$-channel diagrams pick up a relative phase when the full
amplitude is squared.  We define this phase as $\Delta_{Moller}$ and find
it to be given by
\begin{equation}
\Delta_{Moller}=\phi_u-\phi_t={-{\sqrt {ut}}\over {\Lambda_{NC}^2}}
[c_{12}c_\phi-c_{31}s_\phi]\,,
\end{equation}
with the second equality following from Eq.~(15). 
(We define the Mandelstam 
variables as usual: $t,u=-s(1 \mp \cos \theta)/2$.) 
Hence the resulting 
differential distributions for this process appear exactly as in the SM 
except that the $t,u$-channel interference terms should be multiplied by 
$\cos \Delta_{Moller}$. Note that all of the terms involving 
time-space non-commutativity have dropped out of the expression for 
$\Delta_{Moller}$.  In addition, as we take the 
limit $\Lambda_{NC}\to \infty$, $\cos \Delta \to 1$ so that the SM is 
recovered.  In the limit of small $s/\Lambda^2_{NC}$,
$\cos \Delta_{Moller}$ can be expanded where it is seen that
the lowest order correction to the  SM occurs at dimension-8.  This is 
similar to the case of graviton exchange{\cite {virtual}} in  
models with large extra dimensions\cite{ADD} or in the Randall-Sundrum model
of localized gravity{\cite {RS}} below threshold for 
graviton resonance production{\cite {dhr}}.  Perhaps the most 
important thing to notice, as discussed above, is that $\Delta_{Moller} 
\neq 0$ induces a $\phi$ 
dependence in a $2\to 2$ scattering process since there now exists a preferred 
direction in the laboratory frame. 

For simplicity in our numerical results presented below, we will only 
consider the case 
$c_{12}\neq 0$.  If instead $c_{31}$ is non-zero, the results will be  
similar except for the phase of the $\phi$ dependence. When both terms are 
present, the situation is in general somewhat more complex, yet will be
qualitatively comparable 
to the case analysed below. Since we only consider one non-vanishing
value of  $c_{ij}$ at a time, we set its magnitude to unity when obtaining 
our results.

The differential cross section for Moller scattering in the
laboratory center of mass frame can be written as
\begin{equation}
{d\sigma \over {dz~d\phi}}={\alpha^2\over {4s}}\bigg[(e_{ij}+f_{ij})
\big(P^{uu}_{ij}+P^{tt}_{ij}+2P^{ut}_{ij}\cos \Delta_{Moller}\big)+(e_{ij}-
f_{ij})\big({t^2\over {s^2}}P^{uu}_{ij}+{u^2\over {s^2}}P^{tt}_{ij}\big)
\bigg]\,,
\end{equation}
where $z=\cos \theta$, a sum over the gauge boson indices is implied, 
$e_{ij}=(v_iv_j+a_ia_j)^2$ and $f_{ij}=(v_ia_j+a_iv_j)^2$ are combinations of 
the electron's vector and axial vector couplings and 
\begin{equation}
P^{qr}_{ij}=s^2 {{(q-m_i^2)(r-m_j^2)+\Gamma_i \Gamma_j m_i m_j}\over {
[(q-m_i^2)^2+(\Gamma_i m_i)^2][(r-m_j^2)^2+(\Gamma_j m_j)^2]}}\,,
\end{equation}
with $m_i (\Gamma_i)$ being the mass (width) of the $i^{th}$ gauge boson,
where $i$=1(2) corresponds to the photon($Z$). The 
expression for the differential Left-Right Polarization asymmetry, 
$A_{LR}(z,\phi)$, can be easily obtained from the above by forming the ratio
\begin{equation}
A_{LR}(z,\phi)=N(z,\phi)/D(z,\phi)\,,
\end{equation}
where $D(z,\phi)$ is the differential cross section expression above and 
$N(z,\phi)$ can be obtained from $D(z, \phi)$ by the redefinition of 
the coupling combinations $e_{ij}$ and $f_{ij}$ as  
\begin{equation}
e_{ij}=f_{ij}=(v_iv_j+a_ia_j)(v_ia_j+a_iv_j)\,.
\end{equation}
Although we have expressed the cross section in an apparently covariant
form using Mandelstam variables, it is not actually invariant due to
the presence of $\Delta_{Moller}$ which is highly frame dependent.

\vspace*{-0.5cm}
\nn
\begin{figure}[htbp]
\centerline{
\psfig{figure=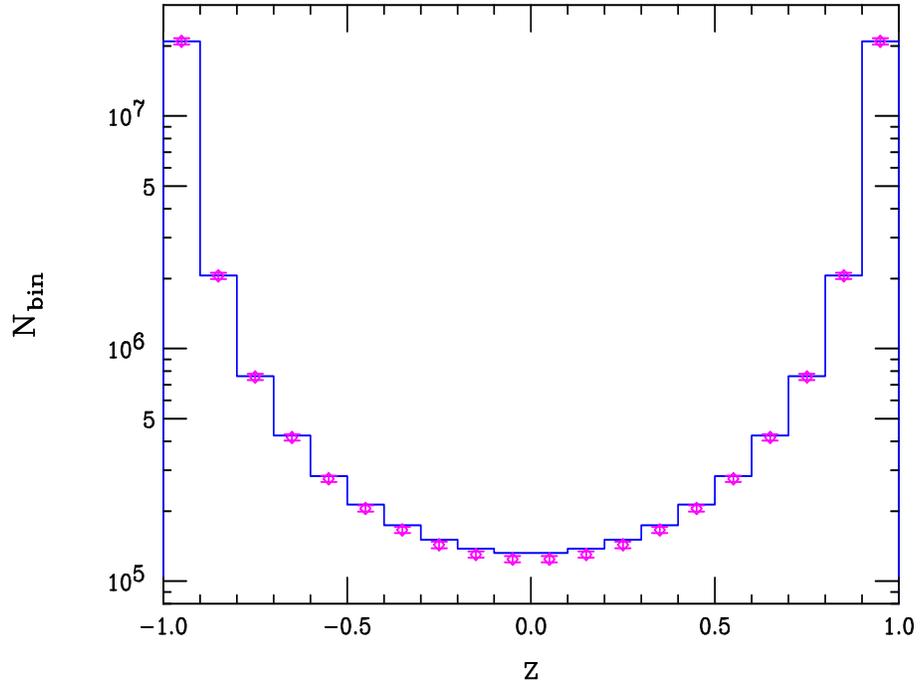,height=9cm,width=12cm,angle=90}}
\vspace*{15mm}
\centerline{
\psfig{figure=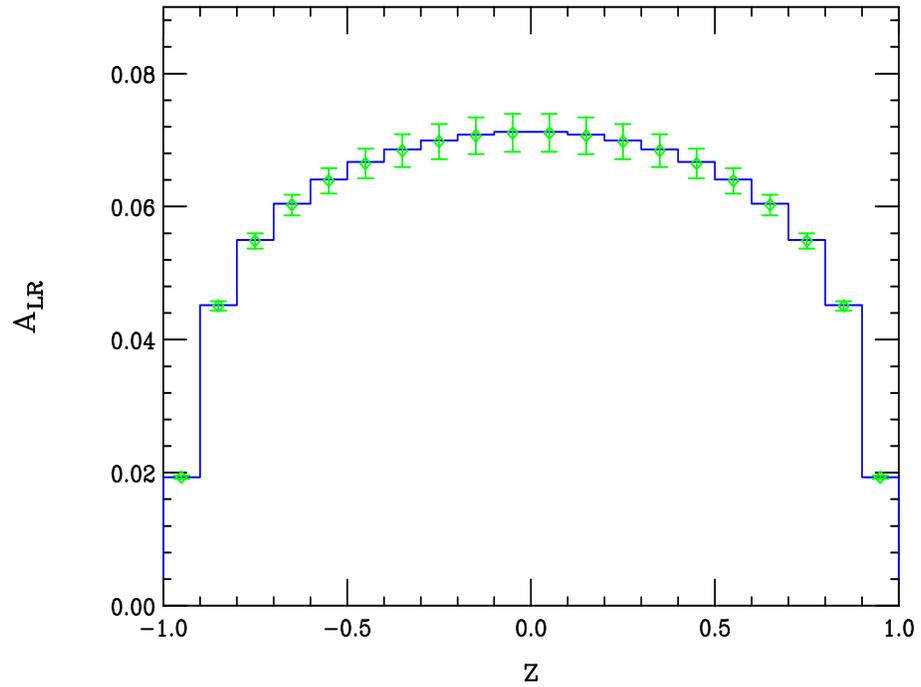,height=9cm,width=12cm,angle=90}}
\caption{Binned cross section (top) and polarized asymmetry (bottom) 
as a function of $z=\cos \theta$ for Moller scattering at a 500 GeV 
linear collider assuming an integrated 
luminosity of 300~fb$^{-1}$. The histogram is the SM expectation while the 
data corresponds to $\Lambda_{NC}=\sqrt s$.}
\label{mollz}
\end{figure}
\vspace*{0.4mm}
\vspace*{-0.5cm}
\nn
\begin{figure}[htbp]
\centerline{
\psfig{figure=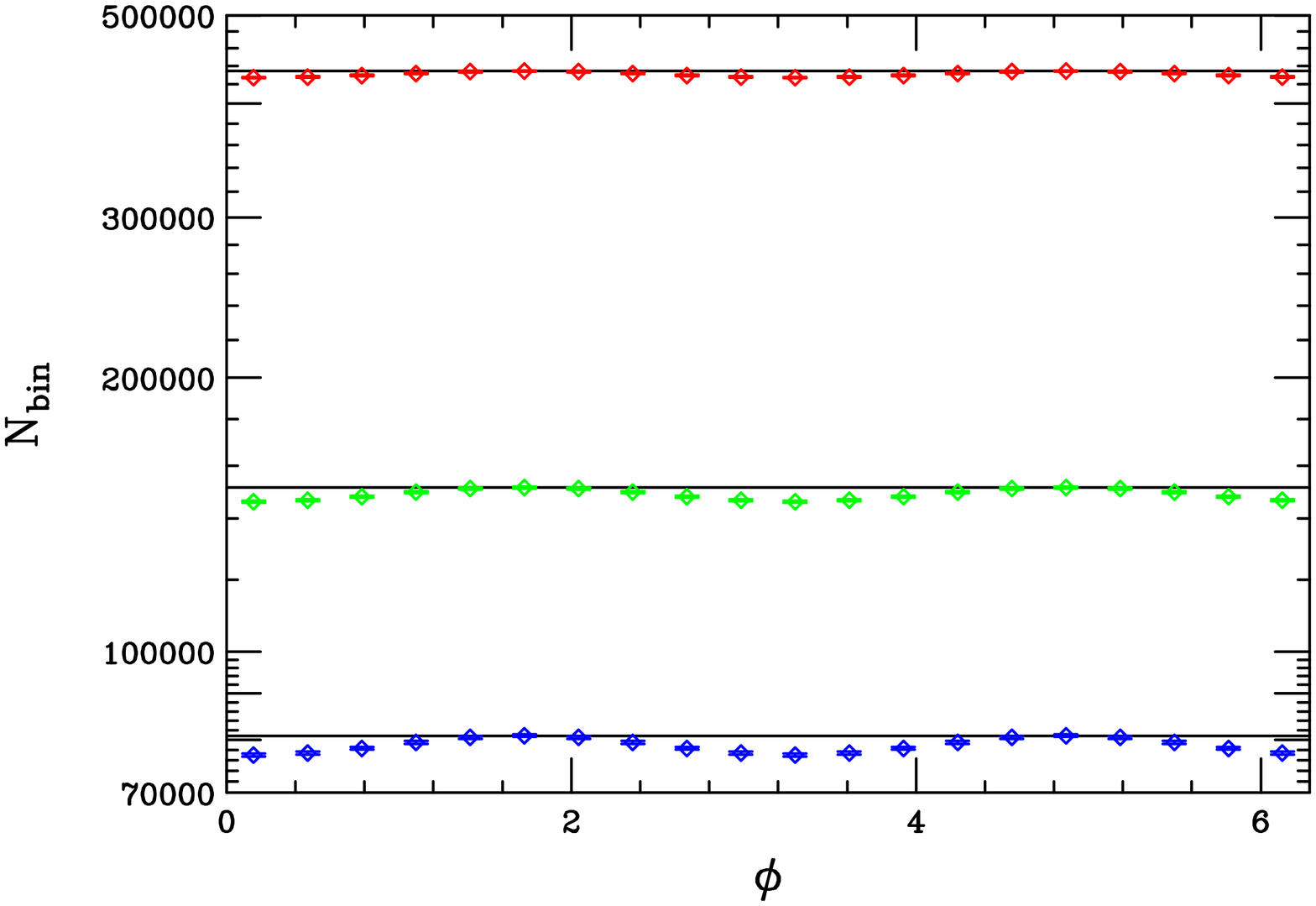,height=9cm,width=12cm,angle=90}}
\vspace*{15mm}
\centerline{
\psfig{figure=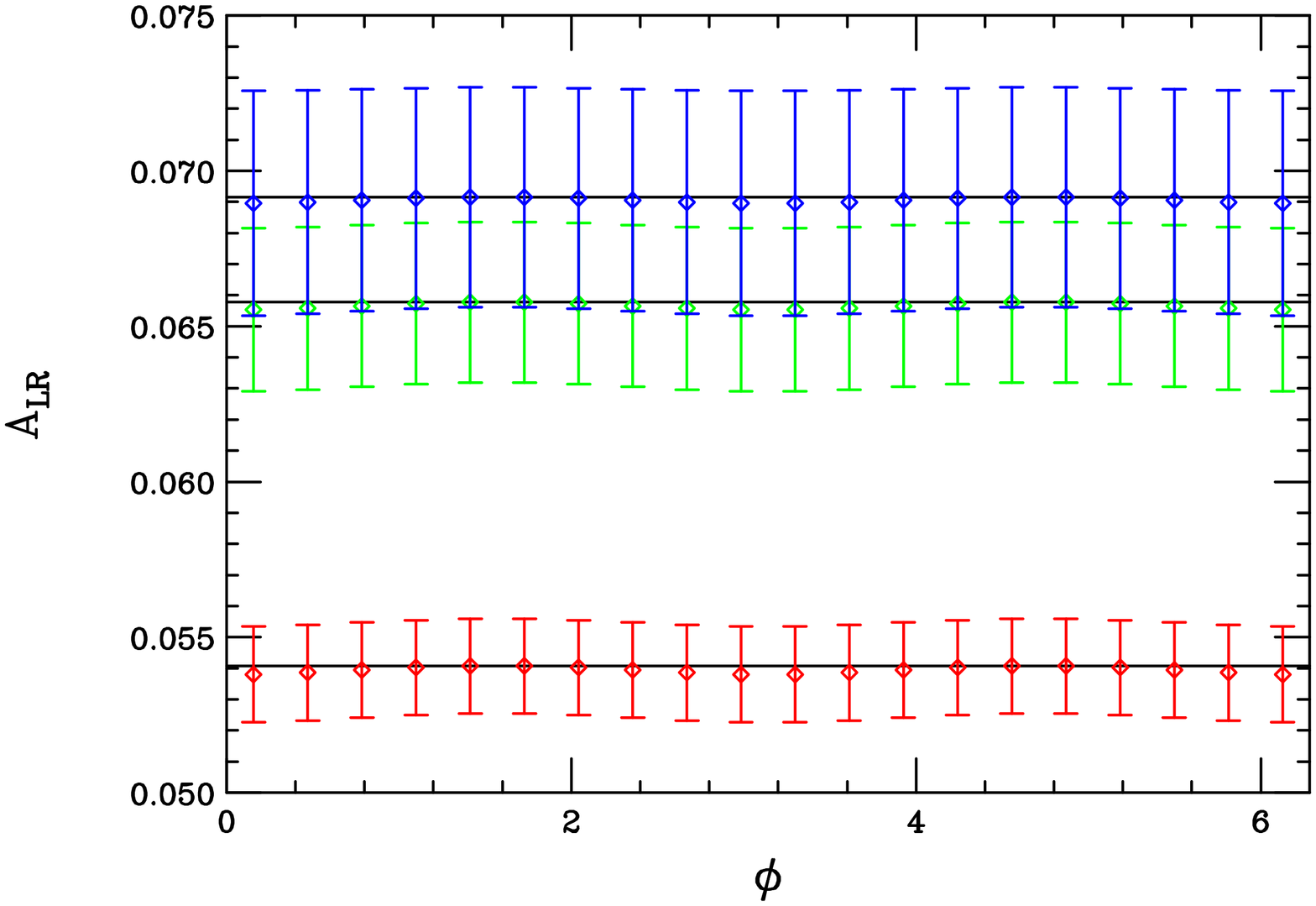,height=9cm,width=12cm,angle=90}}
\caption{$\phi$ dependence of the Moller cross section (top) and left-right
asymmetry (bottom) for the SM (straight lines) and for the case 
$\Lambda_{NC}=\sqrt s$ (shown as data) at a 500 GeV linear collider with 
a luminosity of 300 fb$^{-1}$. From top to 
bottom in the top panel a $z$ cut of 0.9(0.7, 0.5) has been applied with the 
order reversed in the lower panel.}
\label{mollphi}
\end{figure}
\vspace*{0.4mm}

Figure \ref{mollz} displays the effect of a finite value of 
$\Lambda_{NC}=\sqrt s$ on the 
shape of the conventional bin-integrated, $z$-dependent event rate and 
$A_{LR}$ for a 500 GeV linear 
collider assuming an integrated luminosity of 300 $fb^{-1}$.
In presenting these results we have neglected 
initial state radiation and beamstrahlung effects, assumed both beams are 
$90\%$ polarized with $\delta P/P=0.003$, and taken an overall luminosity 
error of 
$1\%$. Angular cuts of $\theta=10^o$ have also been applied but the entire 
$\phi$ range has been integrated over. As we can see from this figure, the 
influence of $\Lambda_{NC}^{-1} \neq 0$ appears to cause a small downward 
shift in the $\cos\theta$ 
distribution which is most noticeable at large scattering 
angles away from the forward and backward $t$- and $u$-channel poles from the 
photon exchange graph. The effect of a finite value of  $\Delta_{Moller}$ 
is thus seen to increase the 
amount of destructive interference between the $u$- and $t$-channel graphs. 
Although the shift is apparently small it occurs over 
many bins and is statistically quite significant given the size of the errors 
at this integrated luminosity. For $A_{LR}$ there is hardly any shift from 
the SM values in this case.

Figure \ref{mollphi} presents the $z$-integrated, $\phi$ dependent 
distribution for 
both the rate and $A_{LR}$. Note that as we perform more restrictive cuts
on $|z|$, the  central region, which is the most sensitive to 
$\Lambda_{NC}$, is becoming more isolated.
As can be seen from the figures, this approach enhances the  
$\phi$ dependence for the differential cross 
section. Though the $\phi$ dependence also appears to be rather weak, 
it is again statistically significant at this large integrated luminosity. As 
in the case of the $\phi$-integrated $A_{LR}$, the $z$-integrated $A_{LR}$ 
shows hardly any sensitivity to finite $\Lambda_{NC}$ even when a strong 
$|z|$ cut is applied. 

\vspace*{-0.5cm}
\nn
\begin{figure}[htbp]
\centerline{
\psfig{figure=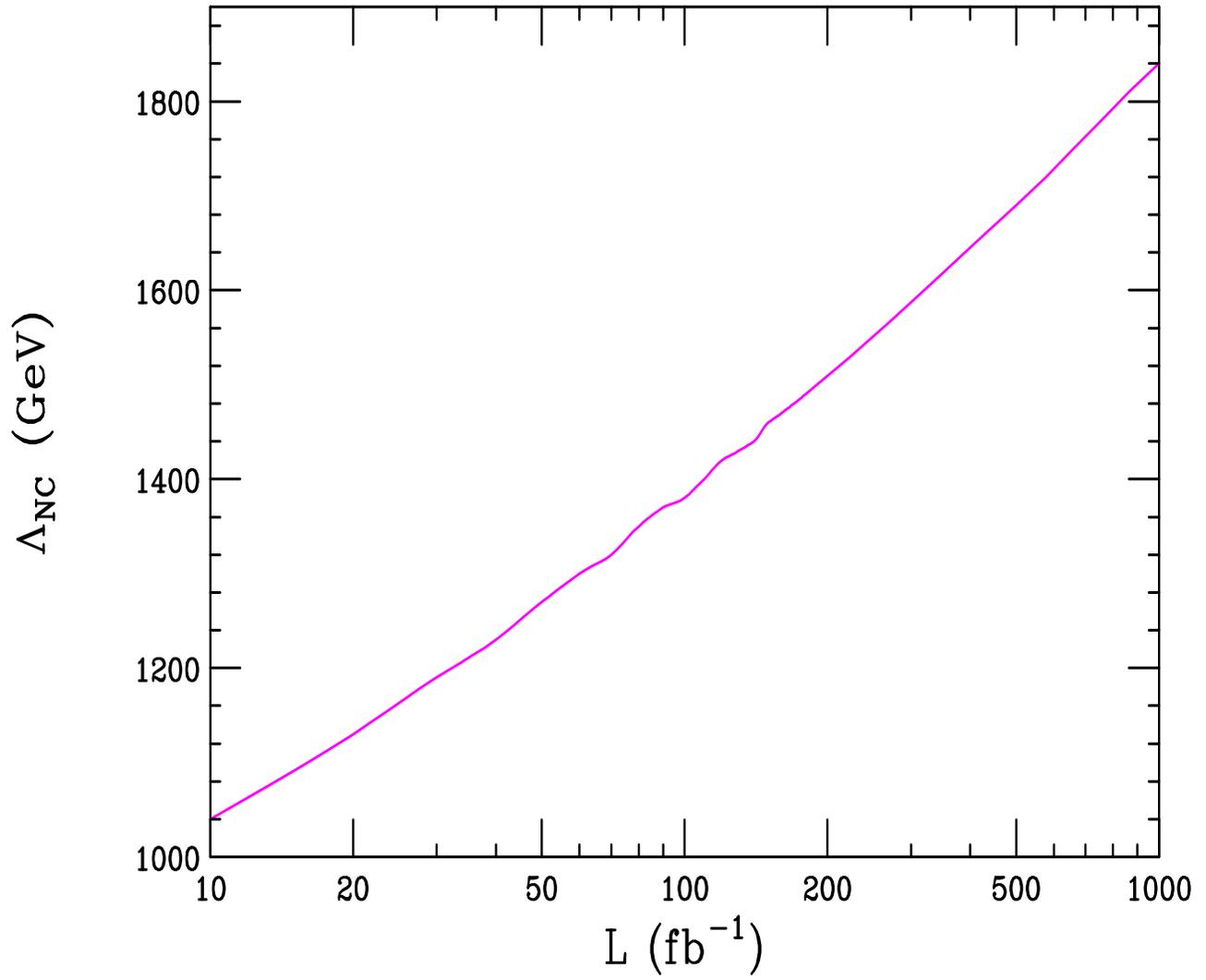,height=14cm,width=17cm,angle=90}}
\vspace*{0.5cm}
\caption[*]{$95\%$ CL lower bound on $\Lambda_{NC}$ at a 500 GeV linear 
collider as a 
function of the integrated luminosity from Moller scattering via the fit 
described in the text.}
\label{molllim}
\end{figure}
\vspace*{0.4mm}

In order to obtain a $95\%$ CL lower bound on $\Lambda_{NC}$ from Moller 
scattering we perform a combined fit to the total cross section, the shape 
of the doubly differential $z-\phi$ angular distribution and $A_{LR}(z,\phi)$.
In the latter two cases we bin the NC results in a $20\times 20$ array of 
$z,\phi$ values and employ only statistical errors apart from the polarization
uncertainty. In the case of the total rate we also include the luminosity 
uncertainty in the error. For 
a fixed value of luminosity we then compare the predictions of the SM with the 
case where $\Lambda_{NC}$ is finite and repeat this procedure by varying 
$\Lambda_{NC}$ until we obtain a $95\%$ CL bound by using a $\chi^2$ fit. 
From this procedure we 
obtain the search reach on $\Lambda_{NC}$ as a function of the integrated 
luminosity as displayed in Fig. \ref{molllim}. As we can see from this 
figure, bounds on $\Lambda_{NC}$ of order $(3-3.5)\sqrt s$ are obtained
for reasonable luminosities.

\vspace*{-0.5cm}
\nn
\begin{figure}[htbp]
\centerline{
\psfig{figure=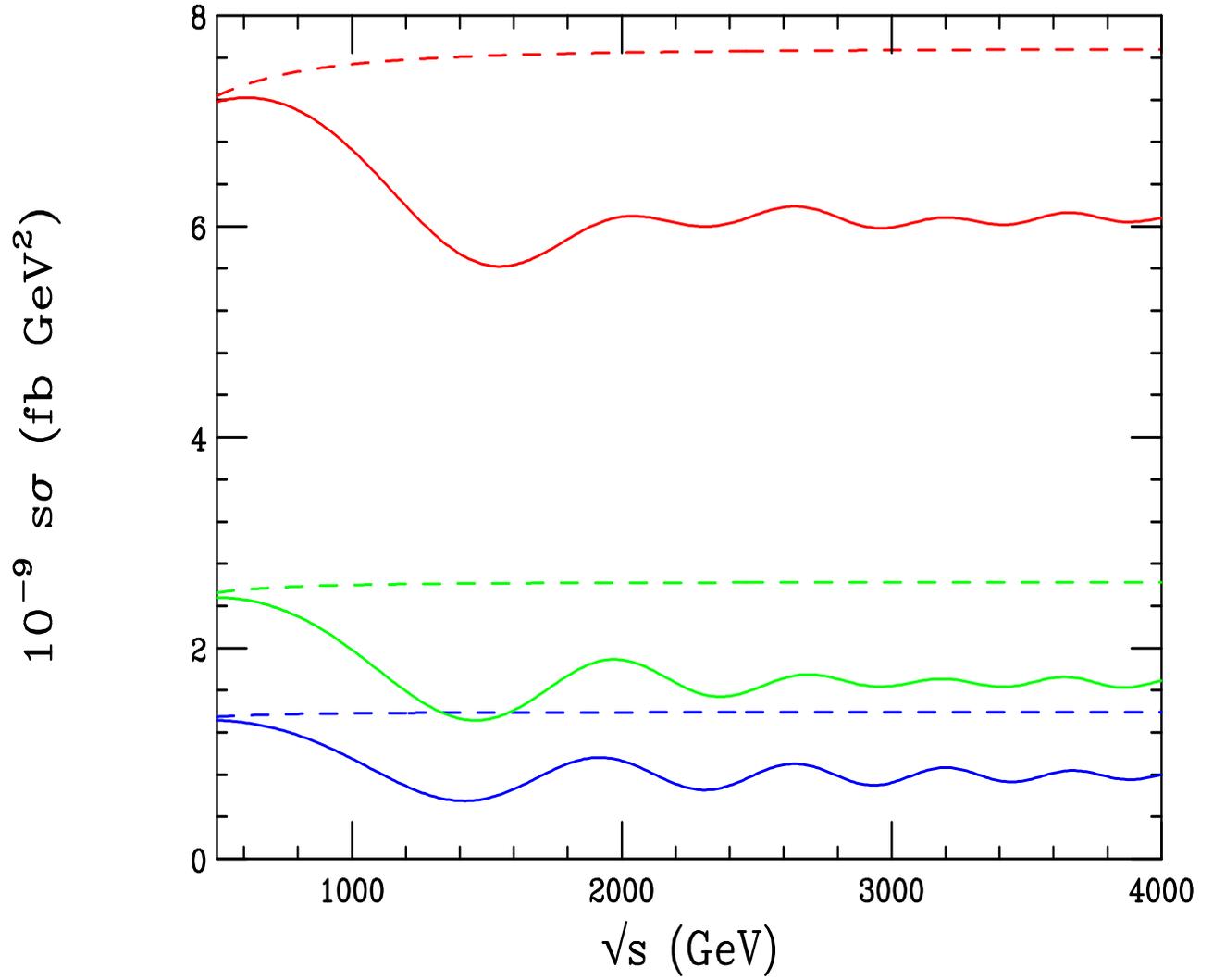,height=14cm,width=17cm,angle=90}}
\vspace*{0.5cm}
\caption[*]{Scaled dependence of the Moller total cross section, subject to a
angular cut (from top to bottom) of  $|z|\leq 0.9(0.7,~0.5)$ assuming the 
SM (dashed curves) or $\Lambda_{NC}=$500 GeV (solid curves).}
\label{mollscaled}
\end{figure}
\vspace*{0.4mm}

We now examine how 
the Moller cross section behaves as $\sqrt s$ grows beyond $\Lambda_{NC}$. 
In the SM for large $s$ we expect the scaled cross section, \ie, the 
product $s\cdot \sigma_{Moll}$, to be 
roughly constant after a cut on $|\cos\theta|$ cut is performed. 
Ordinarily when new operators 
are introduced, the modified scaled cross section is expected to grow
rapidly near the appropriate scale beyond which the contact interaction 
limit no longer applies. However, in the present case,
the theory above the scale $\Lambda_{NC}$ 
{\it is} a well-defined theory since it is not a low energy limit. We would 
thus anticipate that the $\cos \Delta_{Moller}$ factor leads to a modulation 
of the scaled cross section that averages out rapidly  
with a period that depends on the hardness of the $|\cos\theta|$ cut as
the value of $\sqrt s$ increases. This effect is displayed in 
Fig. \ref{mollscaled} and behaves exactly as expected. 

\section{Bhabha Scattering}

\begin{figure}[htbp]
\centerline{
\psfig{figure=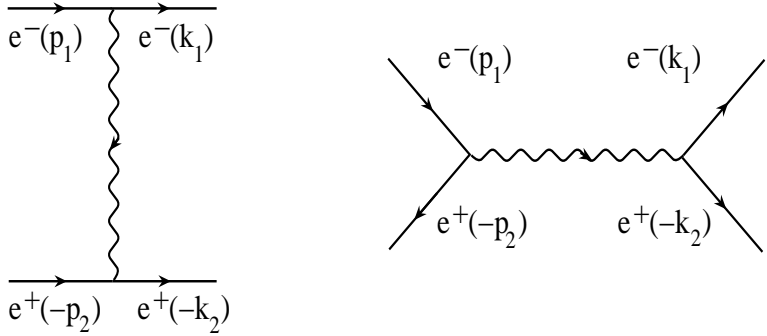,height=6cm,width=12cm,angle=0}}
\vspace*{-0.5cm}
\caption[*]{Feynman graphs contributing to Bhabha scattering with the 
exchanged particle corresponding to a photon and $Z$-boson.} 
\label{bhabha}
\end{figure}

The Feynman graphs which mediate Bhabha scattering in NCQED are given in 
Fig. \ref{bhabha}.  In this case, the 
$t$-and $s$-channel 
kinematic phases are now given by
\begin{eqnarray}
\phi_t &=& {-1\over {2}}[p_1\cdot \theta \cdot k_1-p_2\cdot \theta \cdot 
k_2]\nonumber \\
\phi_s &=& {-1\over {2}}[p_1\cdot \theta \cdot p_2-k_1\cdot \theta \cdot k_2]
\,,
\end{eqnarray}
which implies that the interference term between the two amplitudes is 
sensitive to 
$\cos \Delta_{Bhabha}$ which is given by 
\begin{equation}
\Delta_{Bhabha}=\phi_s-\phi_t={-1\over {\Lambda_{NC}^2}}
[c_{01}t+{\sqrt {ut}}(c_{02}c_\phi+c_{03}s_\phi)]\,.
\end{equation}
Note that whereas Moller 
scattering was sensitive to space-space non-commutativity we see that Bhabha 
scattering is instead sensitive to time-space non-commutativity. Here we see 
that there are two distinct cases depending whether or not $\Delta_{Bhabha}$ 
has a $\phi$ dependence. If $c_{01}$ is non-zero then the $\phi$ dependence 
will be absent, whereas the two cases $c_{02},c_{03}\neq 0$ are essentially 
identical except for the phase of the $\phi$ dependence.  We thus only 
consider the cases $c_{01}=1$ or $c_{02}=1$. 

\vspace*{-0.5cm}
\nn
\vspace*{0.4mm}

\vspace*{-0.5cm}
\nn
\begin{figure}[htbp]
\centerline{
\psfig{figure=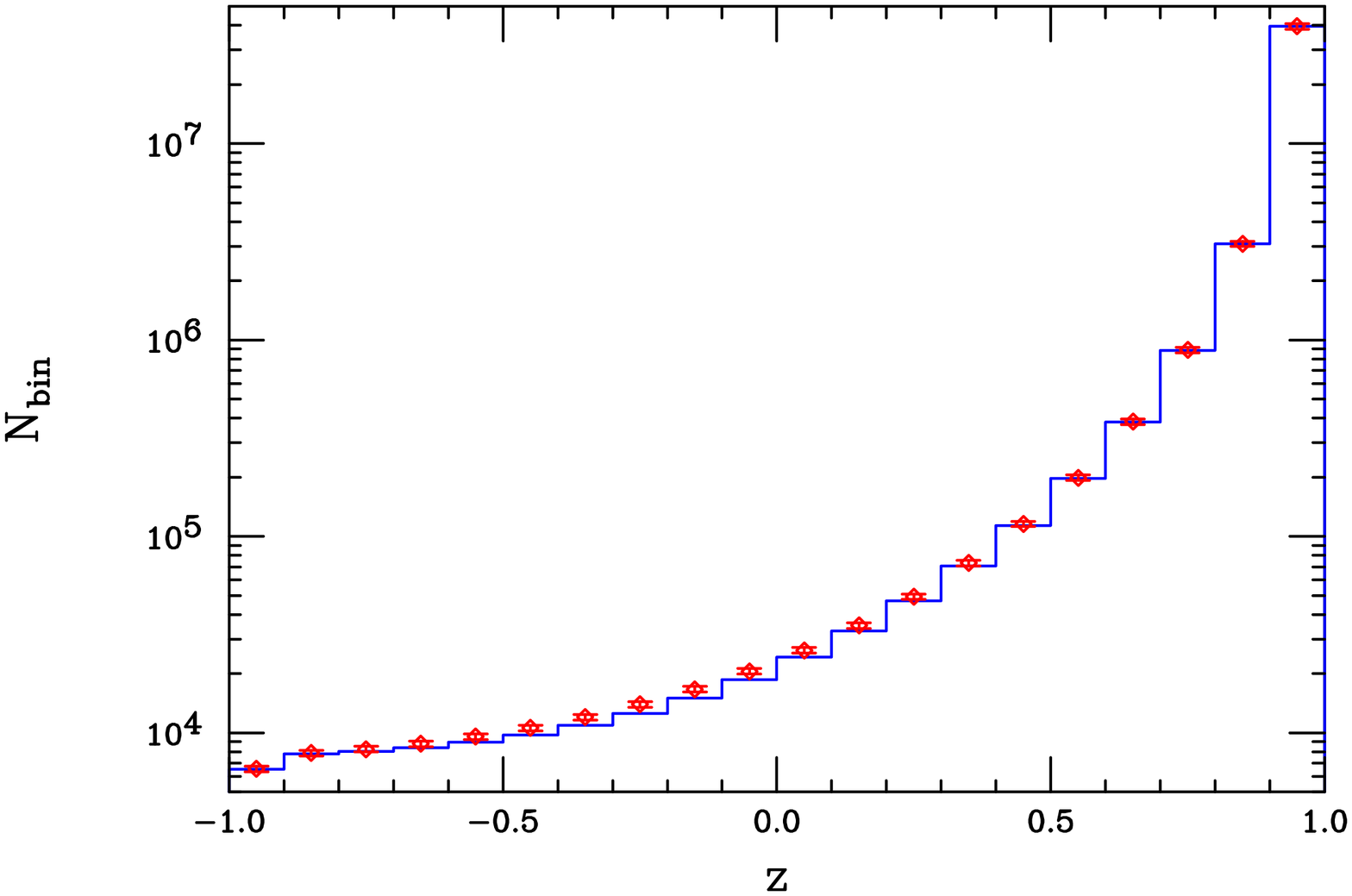,height=9cm,width=12cm,angle=90}}
\vspace*{15mm}
\centerline{
\psfig{figure=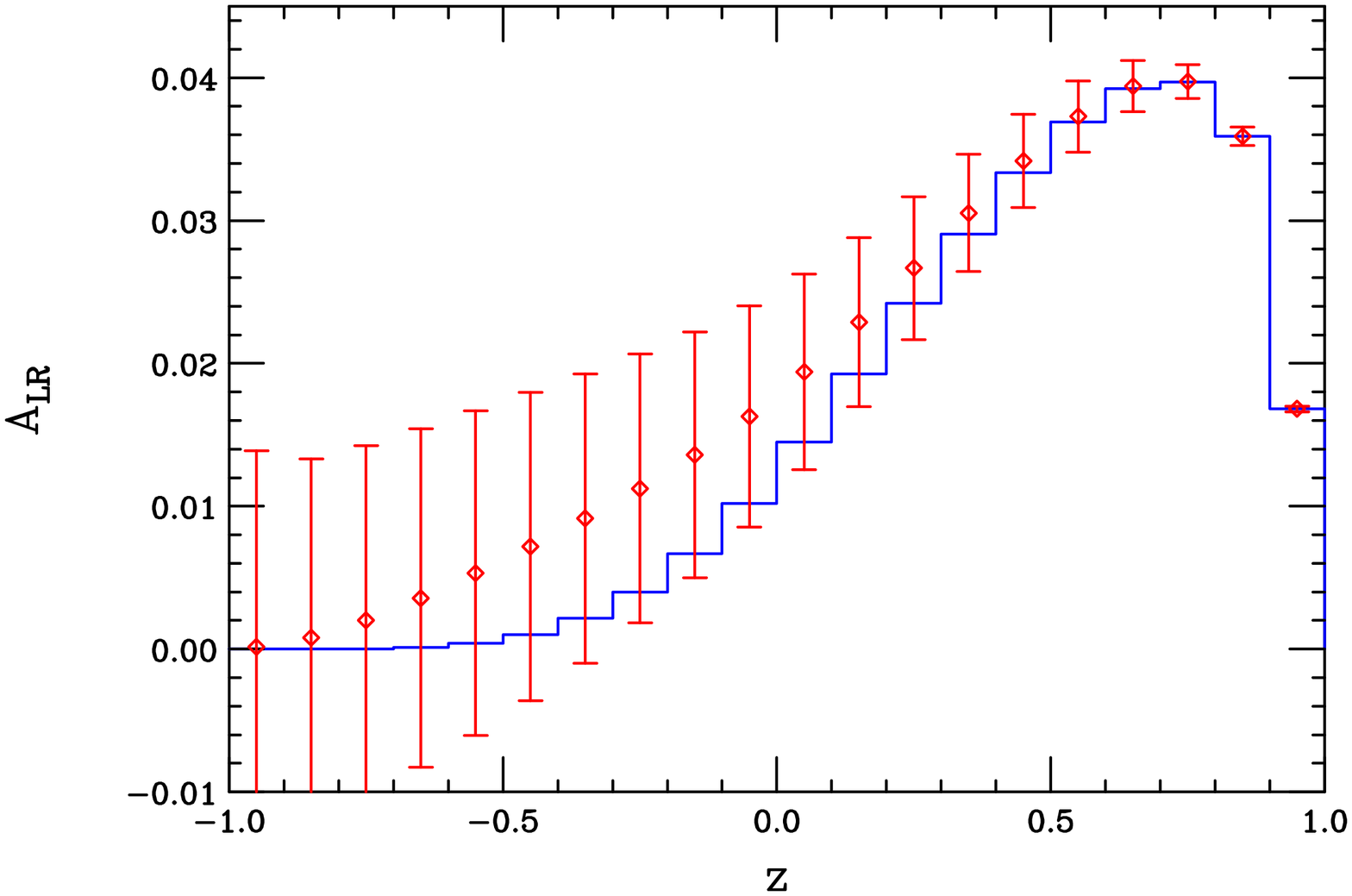,height=9cm,width=12cm,angle=90}}
\caption{Same as Fig. \ref{mollz} but now for Bhabha scattering  assuming 
that $c_{01}$ is non-zero.}
\label{bhazc01}
\end{figure}
\vspace*{0.4mm}

Using the notation above, the differential cross section in the
laboratory center of mass frame for Bhabha 
scattering can then be written as
\begin{equation}
{d\sigma \over {dz~d\phi}}={\alpha^2\over {2s}}\bigg[(e_{ij}+f_{ij})
\big(P^{ss}_{ij}+P^{tt}_{ij}+2P^{st}_{ij} 
\cos \Delta_{Bhabha}\big){u^2\over {s^2}}
+(e_{ij}-f_{ij})\big(P^{ss}_{ij}{t^2\over {s^2}}+P^{tt}_{ij}\big)\bigg]\,,
\end{equation}
with $A_{LR}(z,\phi)$ defined in a manner similar to that for Moller 
scattering by forming the ratio $N(z,\phi)/D(z,\phi)$. 

\vspace*{-0.5cm}
\nn
\begin{figure}[htbp]
\centerline{
\psfig{figure=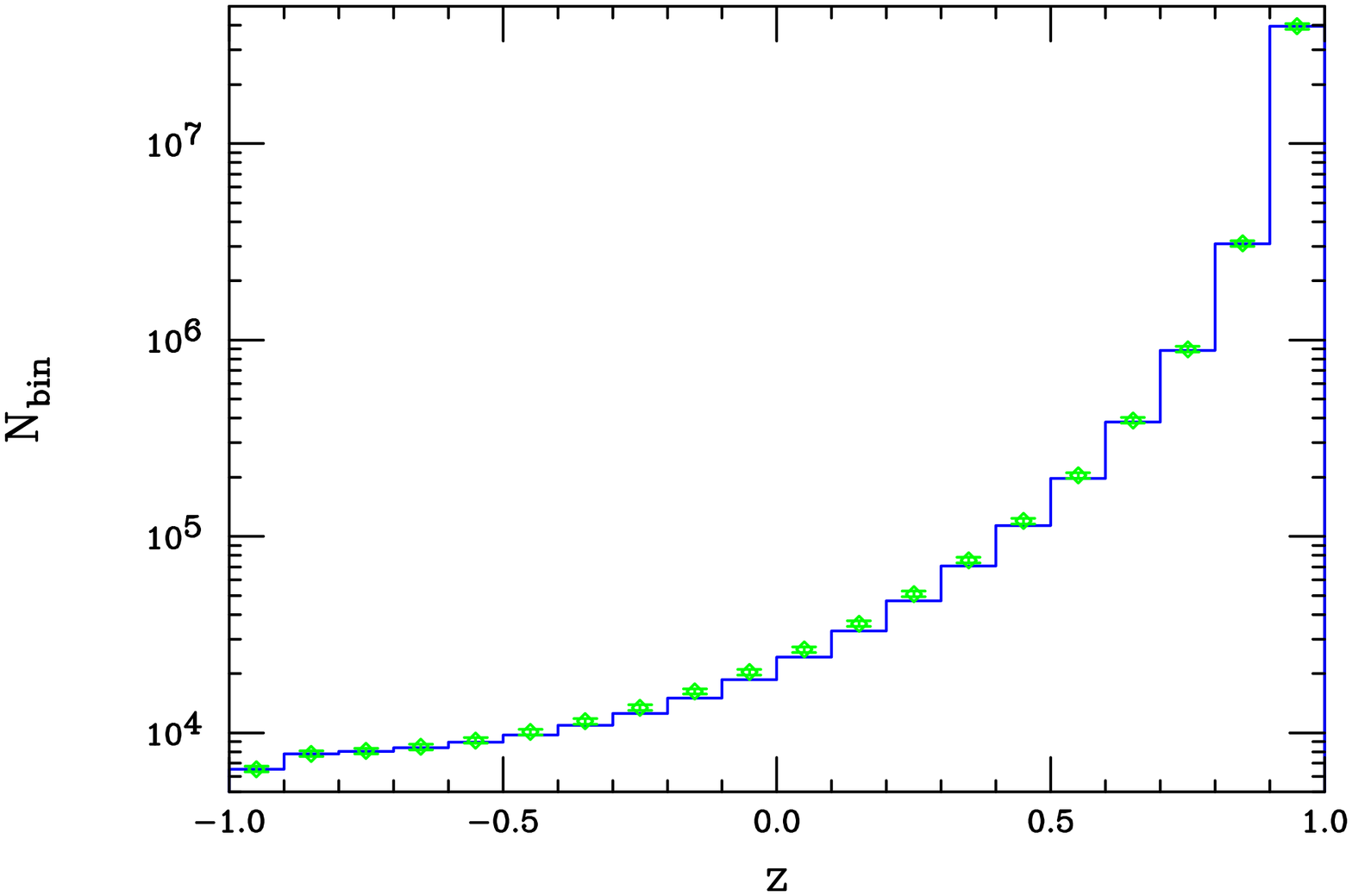,height=9cm,width=12cm,angle=90}}
\vspace*{15mm}
\centerline{
\psfig{figure=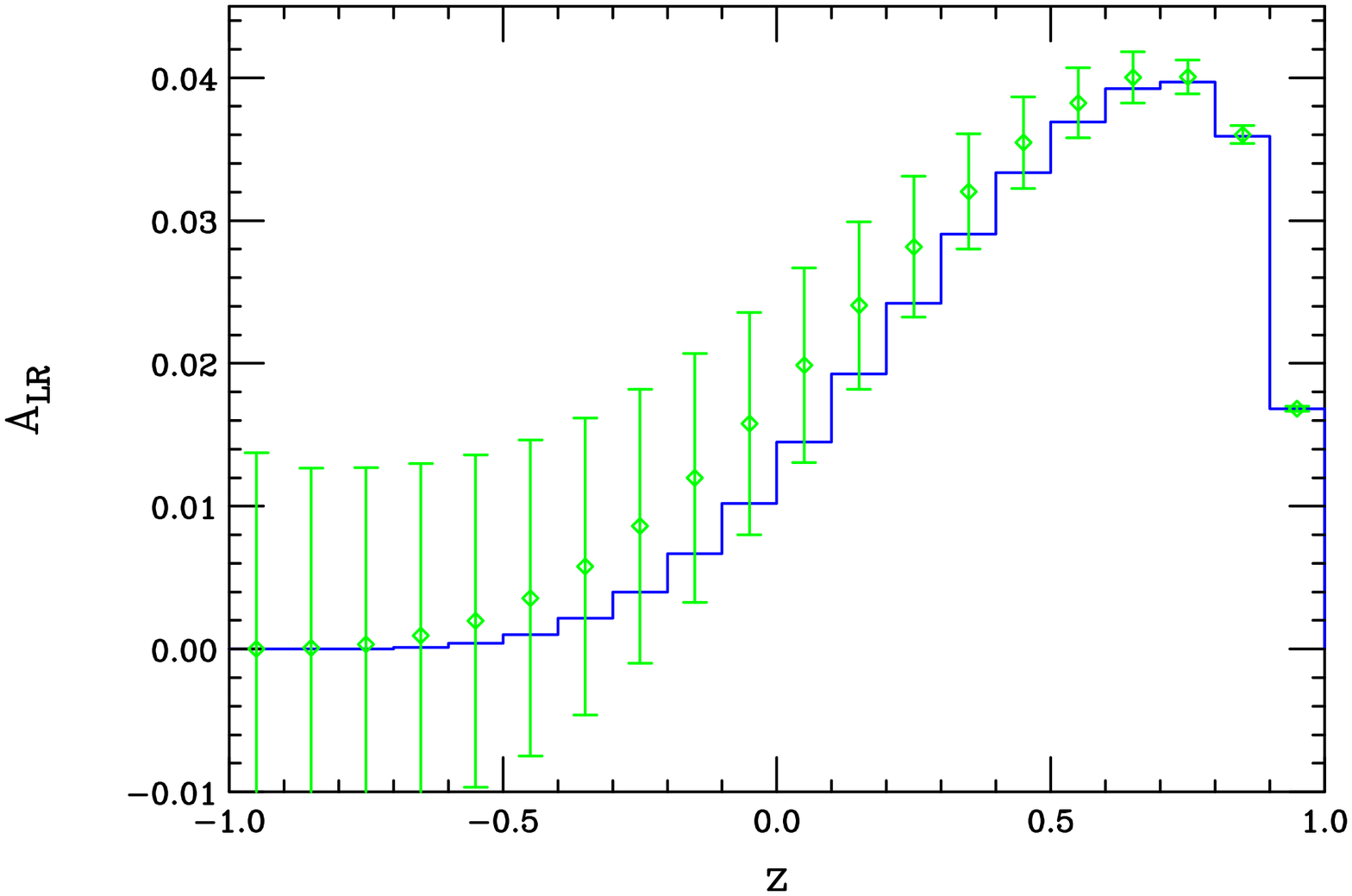,height=9cm,width=12cm,angle=90}}
\caption{Same as in Fig. \ref{bhazc01} but now assuming that $c_{02}$ 
is non-zero.}
\label{bhazc02}
\end{figure}
\vspace*{0.4mm}

\begin{figure}[htbp]
\centerline{
\psfig{figure=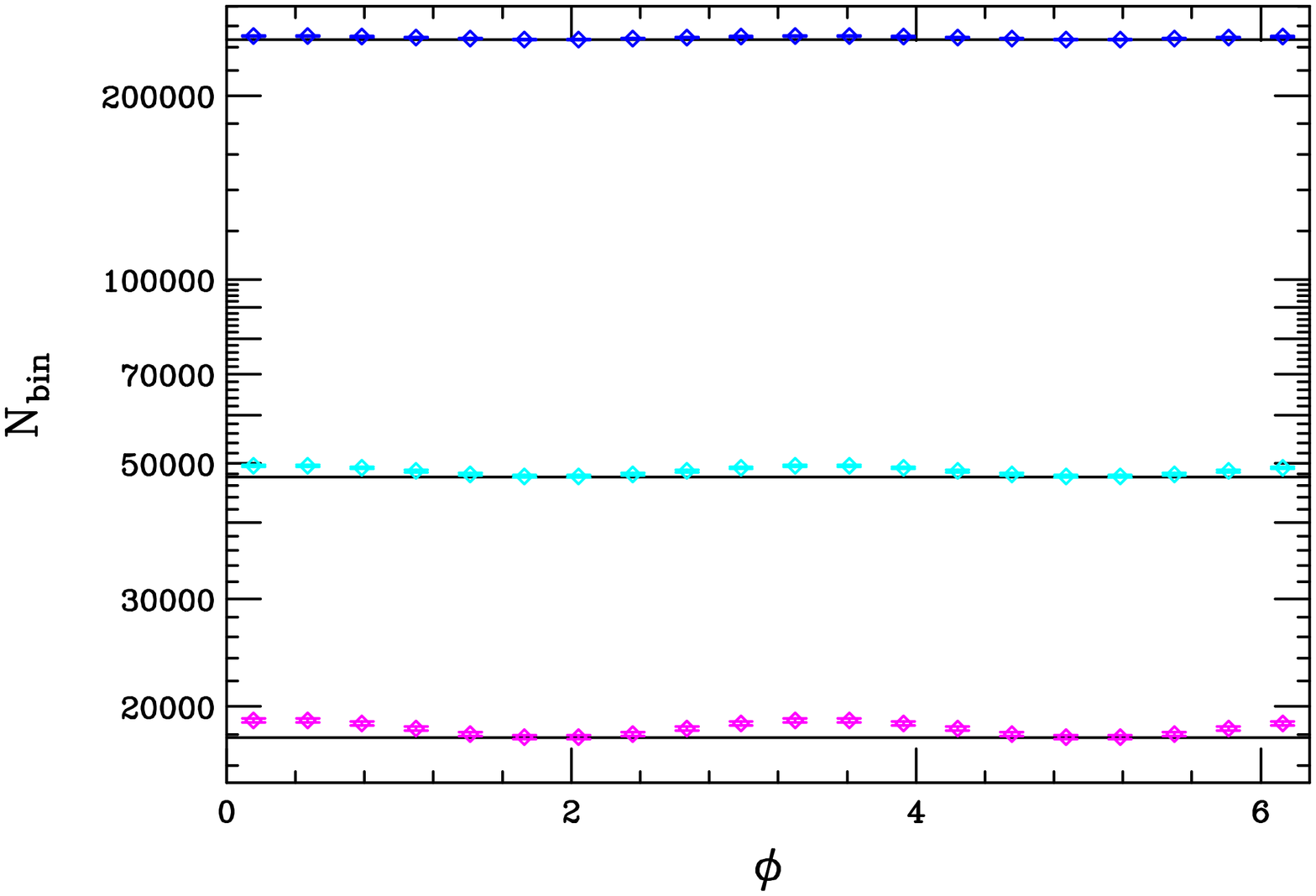,height=9cm,width=12cm,angle=90}}
\vspace*{15mm}
\centerline{
\psfig{figure=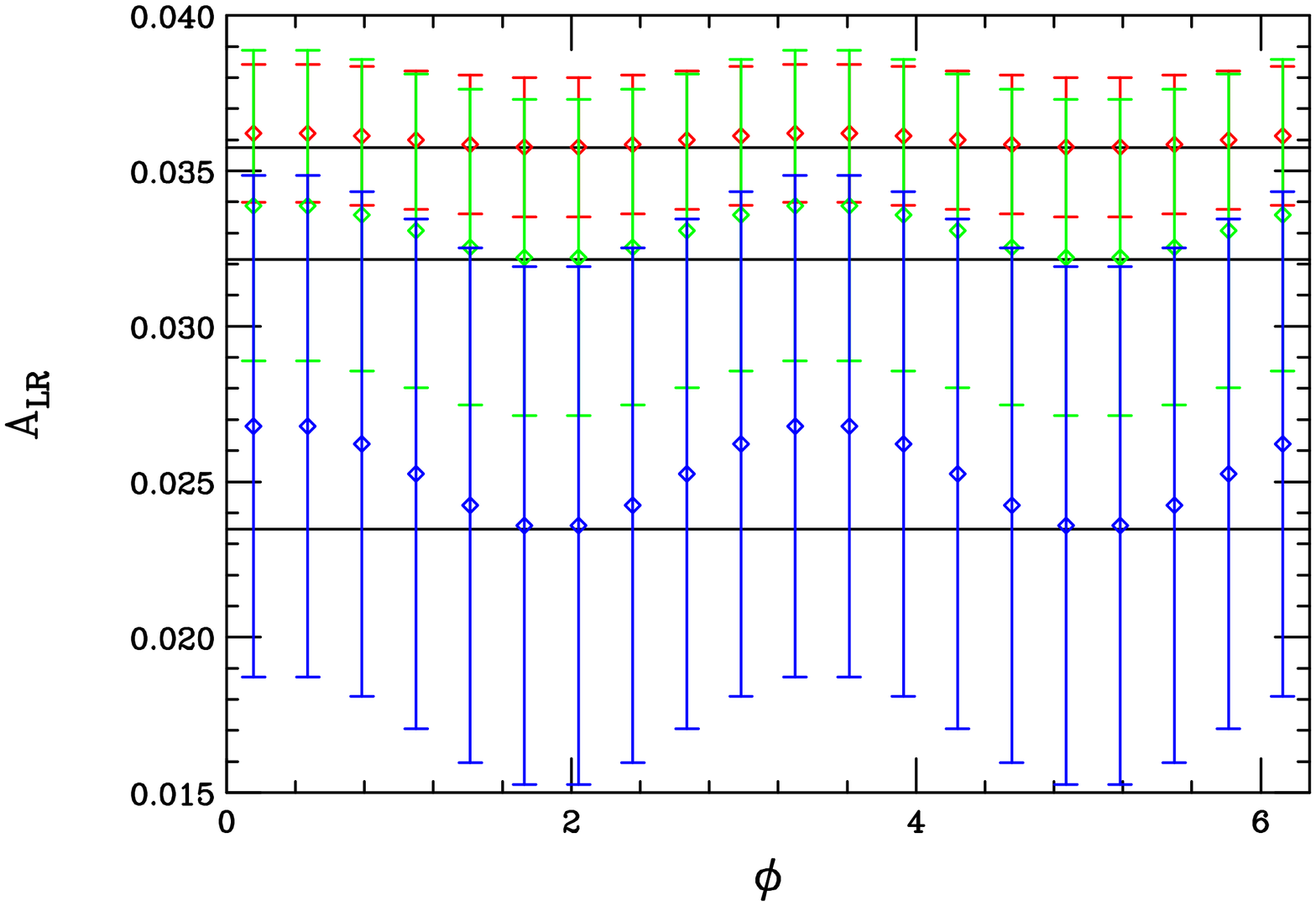,height=9cm,width=12cm,angle=90}}
\caption{Same as Fig. \ref{mollphi} but now for Bhabha scattering with 
$c_{02}$ taken to be non-zero. 
The order for the cuts on $|\cos\theta|$ is reversed in the lower plot.}
\label{bhaphi}
\end{figure}

\begin{figure}[htbp]
\centerline{
\psfig{figure=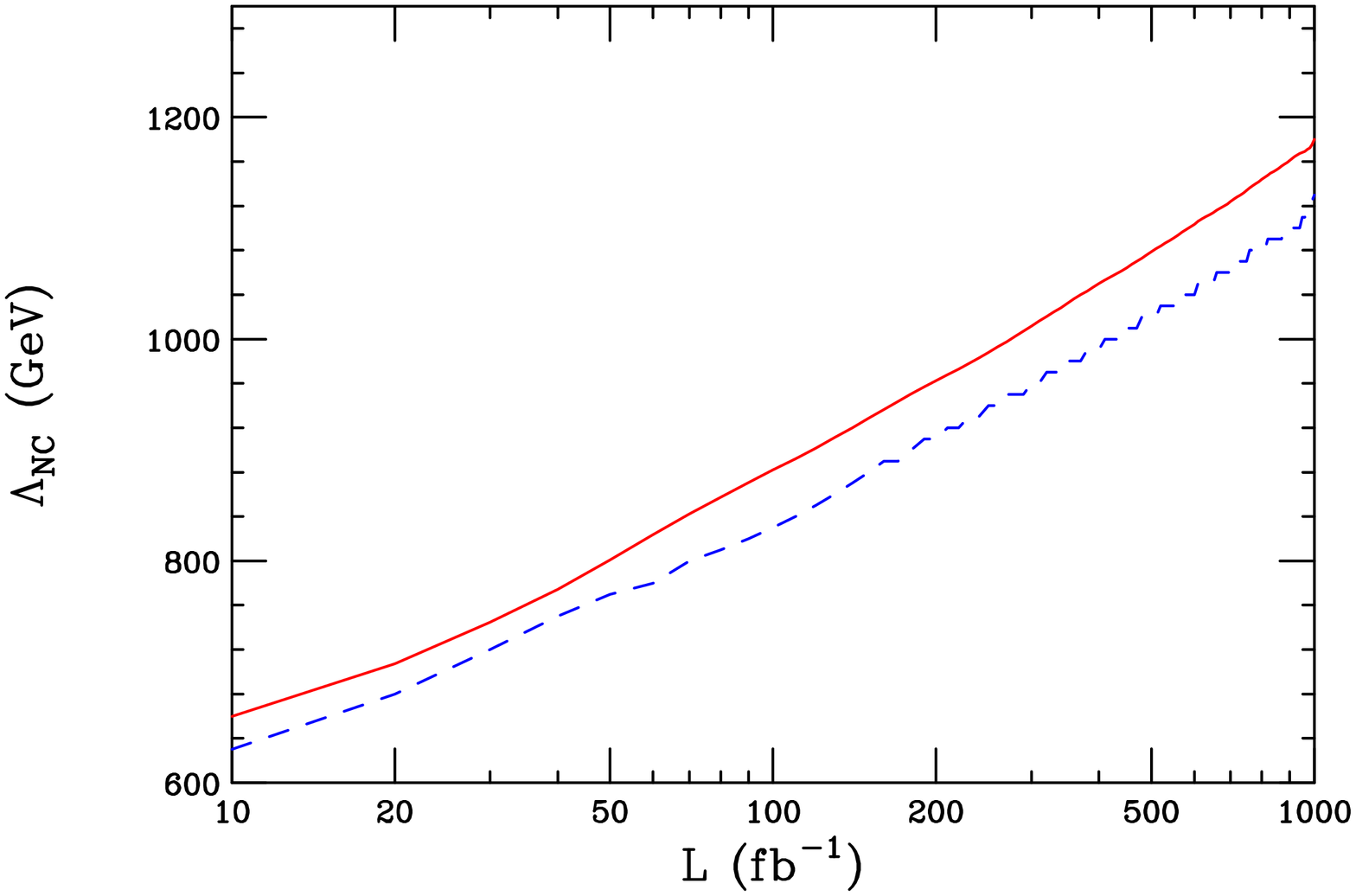,height=14cm,width=17cm,angle=90}}
\vspace*{0.5cm}
\caption[*]{$95\%$ CL bounds on $\Lambda_{NC}$ as a function of luminosity
from Bhabha scattering assuming either 
$c_{01}$ (solid) or $c_{02}$ (dashed) is non-zero.}
\label{bhalim}
\end{figure}

We first consider the case where $c_{01}$ is taken to be non-zero. 
Figure \ref{bhazc01} displays the (in this case trivial) $\phi$ integrated 
angular distribution and $A_{LR}$ for the SM with $\Lambda_{NC}=\sqrt s=500$
GeV. Here one sees that a finite value of $\Lambda_{NC}^{-1}$ 
leads to a slight increase in the cross section at large angles and a 
moderate change in $A_{LR}$ in the same $z$ range. In the case where 
$c_{02}$ is non-zero, Fig. \ref{bhazc02} shows the corresponding distributions.
Note that the 
shift in the cross section looks almost identical in the two cases but 
the deviation in $A_{LR}$ is more shallow in the latter case. 
Figure \ref{bhaphi} shows the 
$\phi$ dependence of the $z$-integrated distributions for the same three 
cuts on $\cos\theta$ discussed above in the case of Moller scattering. 
As before we see that 
the effect in the cross section is most visible for stiffer cuts which isolate 
the central region. In the case of $A_{LR}$ the $\phi$ dependence is too 
small at these integrated luminosities to be visible. In order to obtain a 
$95\%$ CL lower bound on $\Lambda_{NC}$ from Bhabha scattering we follow 
the same procedure as that discussed above for Moller scattering and obtain 
the results presented in Fig. \ref{bhalim}. 
Here we see that the reach for $\Lambda_{NC}$ 
via Bhabha scattering is not quite as good as what we had found earlier for 
the case of Moller scattering, given only by $\simeq 2\sqrt s$, 
for both of the cases considered.

\vspace*{-0.5cm}
\nn
\vspace*{0.4mm}

\begin{figure}[htbp]
\centerline{
\psfig{figure=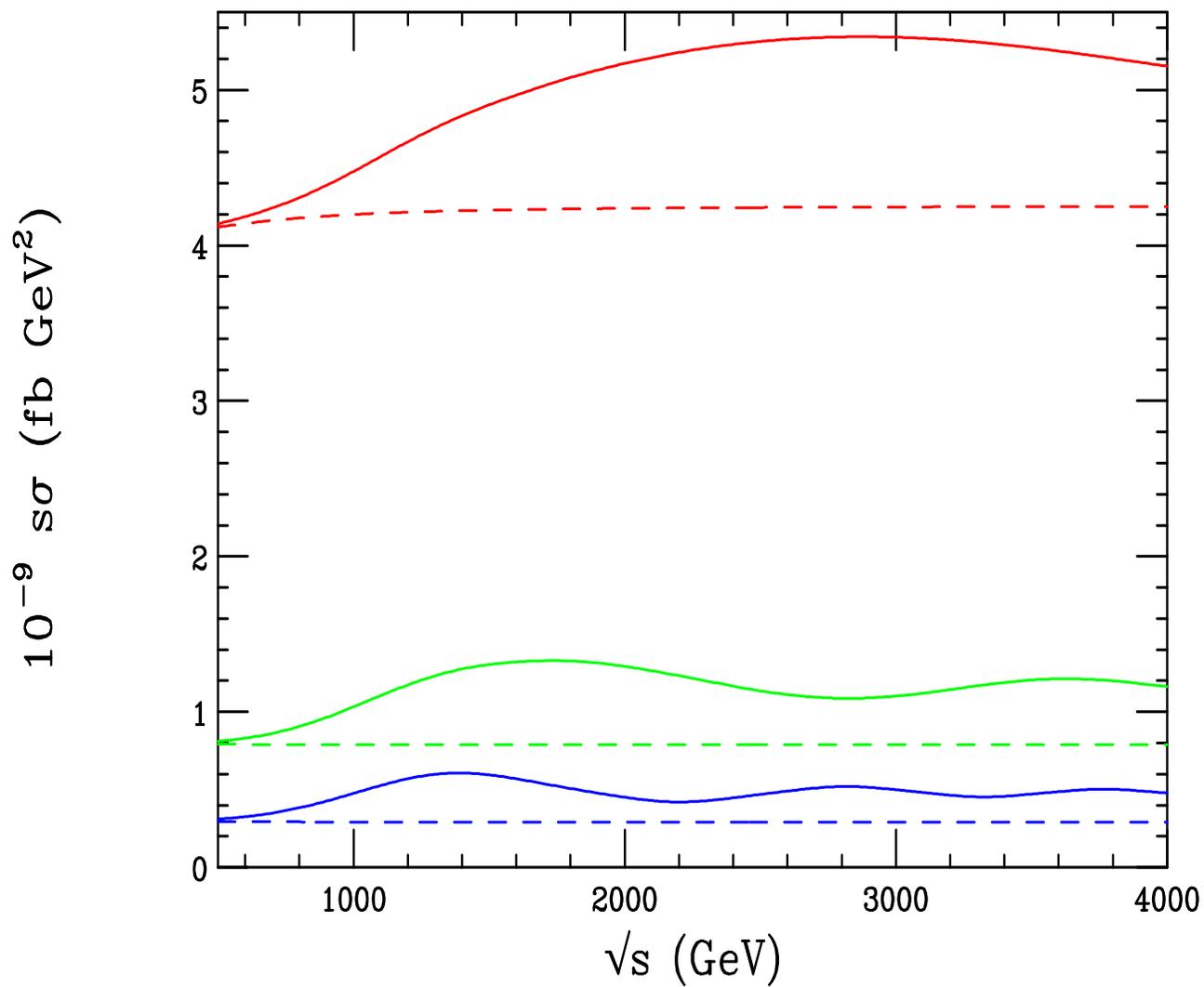,height=14cm,width=17cm,angle=90}}
\vspace*{0.5cm}
\caption[*]{The scaled cross section for Bhabha scattering with $c_{01}$ 
non-zero.}
\label{bscale1}
\end{figure}

\begin{figure}[htbp]
\centerline{
\psfig{figure=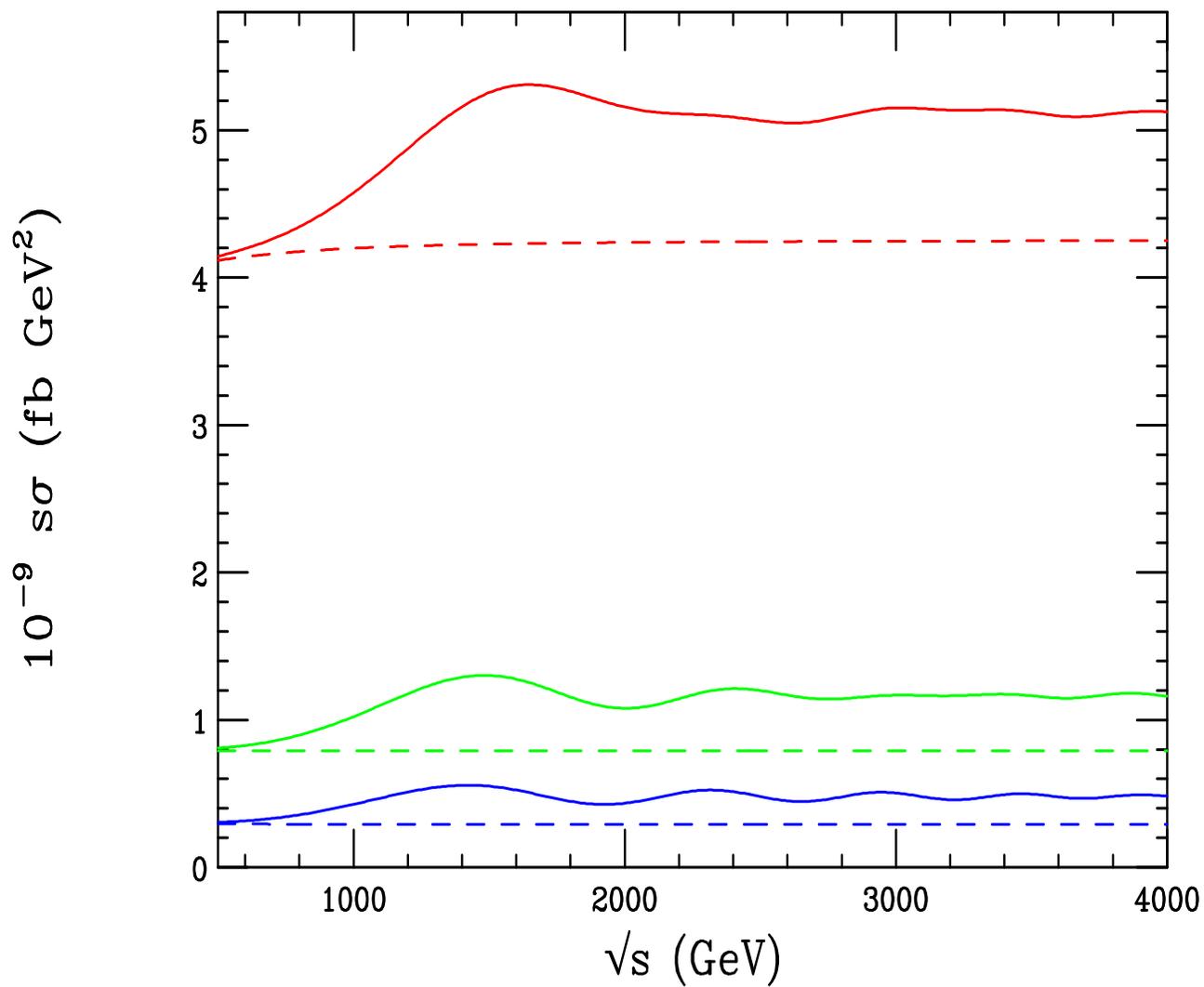,height=14cm,width=17cm,angle=90}}
\vspace*{0.5cm}
\caption[*]{Same as in the previous figure but now with $c_{02}$ non-zero.}
\label{bscale2}
\end{figure}

Figures \ref{bscale1} and \ref{bscale2} show the scaled cross sections 
for Bhabha scattering after the $z$ cuts are employed for values of 
$\sqrt s > \Lambda_{NC}$. Here we 
see that for both cases, the presence of a finite value for 
$\Delta_{Bhabha}$ leads to 
an increase in the constructive interference between the $s$- and $t$-channel 
exchanges with two very different periods. Again for values of $\sqrt s$ 
much larger than $\Lambda_{NC}$ we see that the oscillations average out to 
approximately half of their original amplitude.

\vspace*{-0.5cm}
\nn
\vspace*{0.4mm}
\vspace*{-0.5cm}
\nn
\vspace*{0.4mm}
\vspace*{-0.5cm}
\nn
\vspace*{0.4mm}

\section{Pair Annihilation}

\begin{figure}[htbp]
\centerline{
\psfig{figure=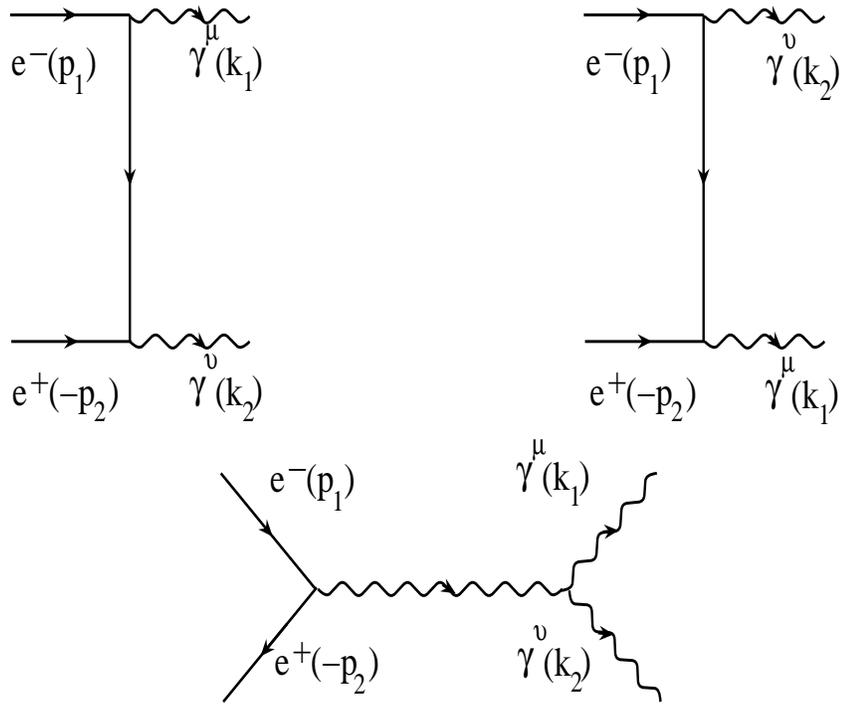,height=10.5cm,width=14cm,angle=0}}
\vspace*{0.5cm}
\caption{The three tree level contributions to $e^+ e^- \rightarrow \gamma 
\gamma$ in NCQED.}
\label{feynpair}
\end{figure}

The Feynman diagrams which contribute to pair annihilation in NCQED are
shown in Fig. \ref{feynpair}.  Note that in this case, there is a novel
s-channel contribution in NC field theories from the $ 3 \gamma$ 
self-coupling,
in addition to the kinematical phase factor which appears at each vertex.
Due to the presence of the non-abelian like coupling, one must exercise
caution in calculating the cross section to ensure that the Ward identities
are satisfied and to guarantee that unphysical polarization states are not
produced.  Hence one must either extend the polarization sum to incorporate
transverse photon
polarization states or include the contribution from the production of
a ghost-antighost pair to cancel the contribution of the unphysical
gauge boson polarizations.  This procedure is similar in manner to that
performed for the parton-level scattering of $q\bar q\to gg$ in QCD.
We find that the differential cross section in the
laboratory center of mass frame for pair annihilation in NCQED 
is then given by
\begin{equation}
\frac{d \sigma}{dz \, d\phi} = \frac{\alpha^2}{4s} \, \bigg[ \frac{u}{t} + 
\frac{t}{u} -4 \, \frac{t^2 +u^2}{s^2} \, {\rm sin}^2 (\frac{1}{2} k_1 \wedge 
k_2) \bigg]\,,
\end{equation}
where we have introduced the wedge product defined as
$p \wedge k = p_\mu k_\nu 
\theta^{\mu \nu}$.  Note that in this case, the contribution from the
relative phases from the interference terms cancels.  We also note
that the sign of the modification due to NCQED does not vary since it
is an even function and hence the effect does not wash out over time
due to the rotation of the Earth.

\noindent

Evaluating the wedge product yields
\begin{equation}
\Delta_{PA}\equiv {1\over2} k_1 \wedge k_2= \frac{-s}{2 \Lambda^2_{NC}} 
\bigg[ c_{01} c_\theta + c_{02} s_\theta c_\phi + c_{03} 
s_\theta s_\phi  \bigg]\,.
\end{equation}
Note that this process is sensitive only to space-time non-commutativity.
We again stress that this is only true in the 
CM frame; due to the violation of Lorentz invariance this will not hold in
all reference frames.  As discussed above,   
it is important to remember that although we have expressed the
cross section in terms of the Lorentz invariant Mandelstam variables,
$s\,, t\,, u$, the phase $\Delta_{PA}$ is not Lorentz invariant. 
For this reaction, we parameterize the $c_{0i}$ by introducing the angles 
characterizing the background {\bf E} field of the theory:
\begin{eqnarray}
c_{01} &=& {\rm cos} \alpha \nonumber \\
c_{02} &=& {\rm sin} \alpha \, {\rm cos} \beta  \\ 
c_{03} &=& {\rm sin} \alpha \, {\rm sin} \beta \nonumber \,, 
\end{eqnarray}
so that
\begin{eqnarray}
\Delta_{PA}&=& \frac{-s}{2 \Lambda^2_{NC}} 
\bigg[ {\rm cos}  \theta \, {\rm cos}
\alpha + {\rm sin} \theta \, {\rm sin} \alpha \, {\rm cos}(\phi -\beta) 
\bigg] \nonumber \\ &=& \frac{-s}{2\Lambda^2_{NC}} \, {\rm cos}\theta_{NC}\,,
\end{eqnarray}
where $\theta_{NC}$ is the angle between the {\bf E} field and the direction 
of the outgoing photon denoted with momenta $k_1$.  
Note that $\beta$ simply defines the origin of the 
$\phi$ axis; we will hereafter set $\beta = \pi /2$.  This parameterization
provides a good physical interpretation of the NC effects.  
(Note that the $c_{0i}$ 
are not independent; in pulling out the overall scale $\Lambda_{NC}$ we
can always impose 
the constraint $ |c_{01}|^2 + |c_{02}|^2 + |c_{03}|^2 =1$.)
Here, we consider three physical cases: $\alpha = 0$, $\alpha = \pi /2$, 
and $\alpha = \pi /4$, which correspond to the background {\bf E} 
fields being at an angle $\alpha$ from 
the beam axis.  The correction term $\Delta_{PA}$ then takes the following 
forms in each of these cases:
\begin{eqnarray}
\Delta_{PA}(\alpha =0) &=& \frac{-s}{2 \Lambda^2_{NC}} \, {\rm cos}\theta 
\nonumber \\
\Delta_{PA}(\alpha = \pi /2) &=& \frac{-s}{2 \Lambda^2_{NC}} \, 
{\rm sin}\theta 
\, {\rm sin}\phi \nonumber \\
\Delta_{PA}(\alpha = \pi /4) &=& \frac{-s}{2 {\sqrt 2} \Lambda_{NC}^2} \, 
\bigg[ {\rm cos}\theta + {\rm sin}\theta \, {\rm sin}\phi \bigg].
\end{eqnarray}
As in the previous processes we considered, a
striking feature of these correction terms are their $\phi$ dependence, 
arising from a preferred direction which is not parallel to the beam axis.  

\vspace*{-0.5cm}
\noindent
\begin{figure}[htbp]
\centerline{
\psfig{figure=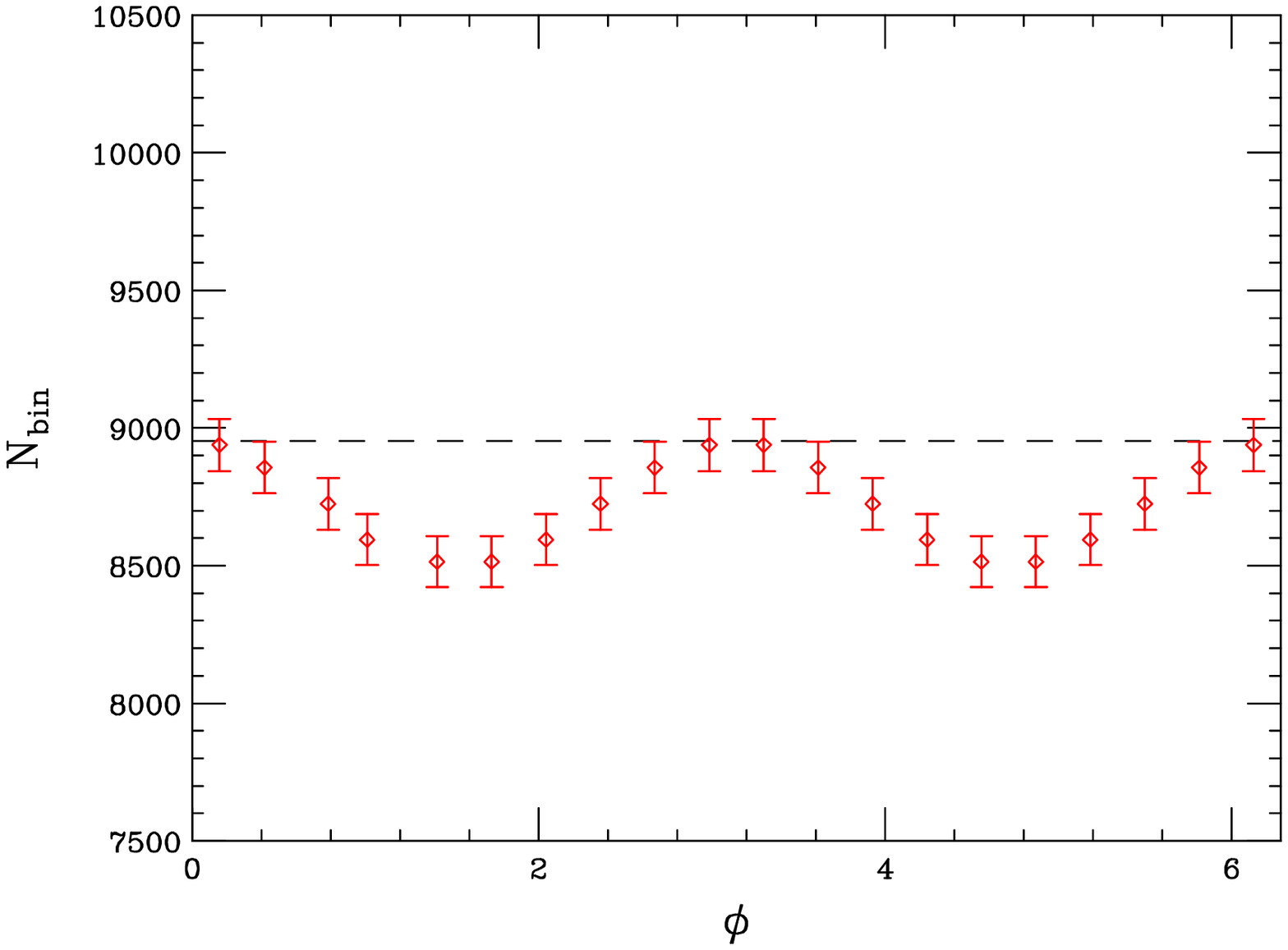,height=9cm,width=12cm,angle=0}}
\vspace*{15mm}
\centerline{
\psfig{figure=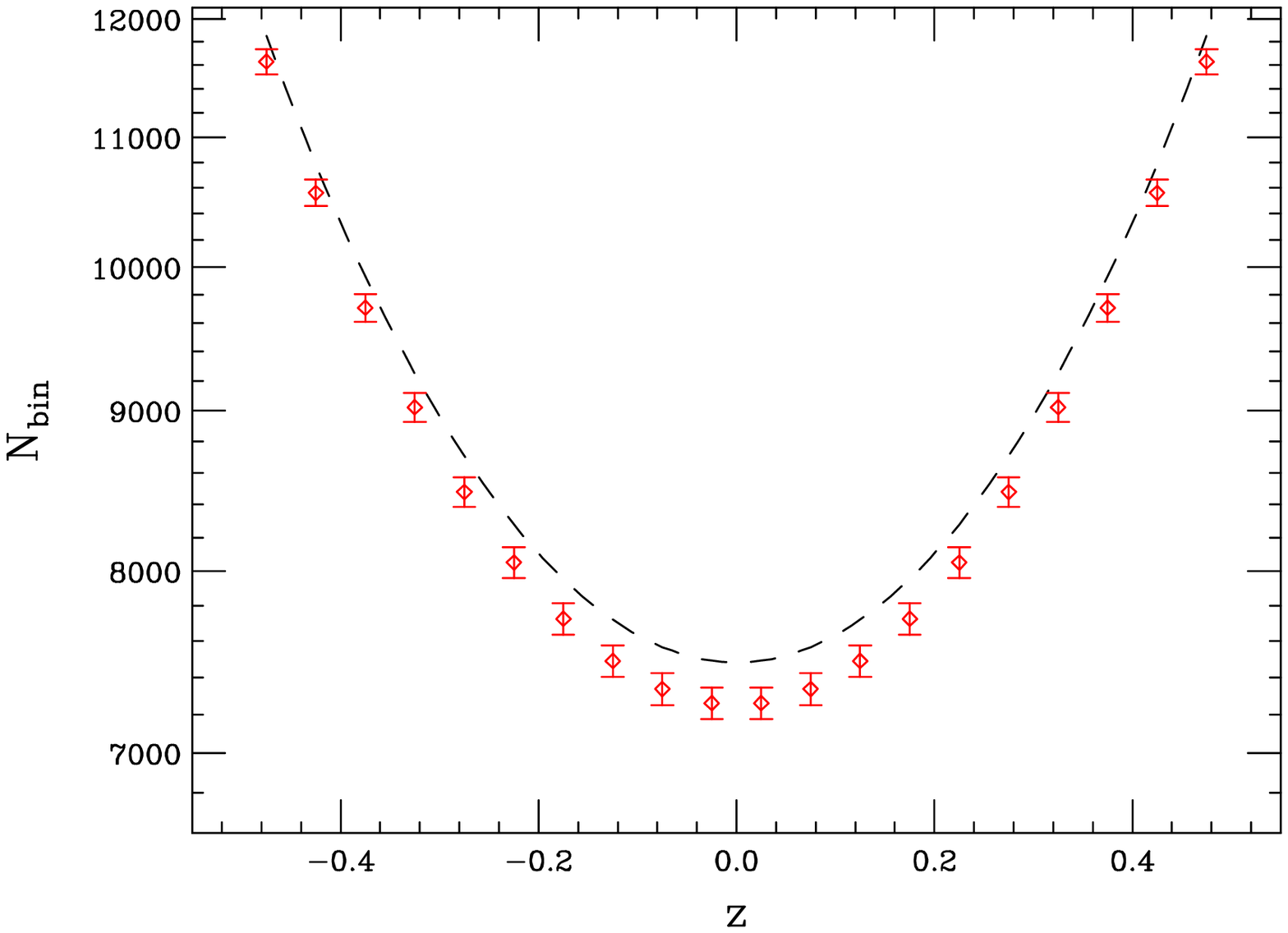,height=9cm,width=12cm,angle=0}}
\caption{$\phi$ dependence (top) and $\theta$ dependence (bottom) of the 
$e^+ e^- \rightarrow \gamma \gamma$ cross section for the case $\alpha = \pi 
/2$.  We take $\Lambda_{NC} = {\sqrt s}= 500$ GeV, and assume a luminosity 
of $500 \, {\rm fb}^{-1}$.  In the top panel a cut of 
$|z| < 0.5$ has been employed.  The dashed line corresponds to the SM 
expectations and the `data' points represent the NCQED results.}
\label{eeggpi2}
\end{figure}
\vspace*{0.4mm}

\vspace*{-0.5cm}
\noindent
\begin{figure}[htbp]
\centerline{
\psfig{figure=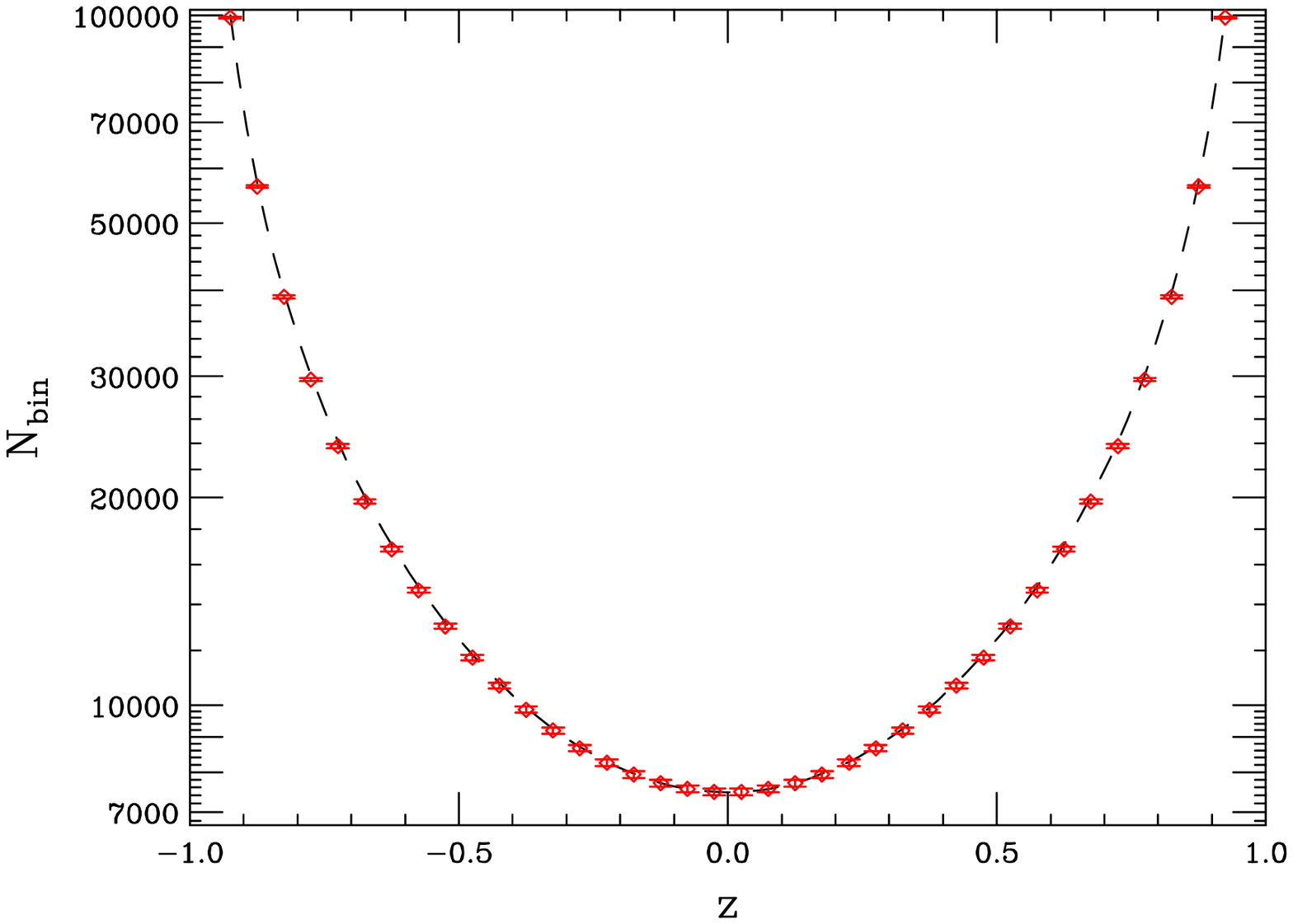,height=9cm,width=12cm,angle=0}}
\vspace*{15mm}
\centerline{
\psfig{figure=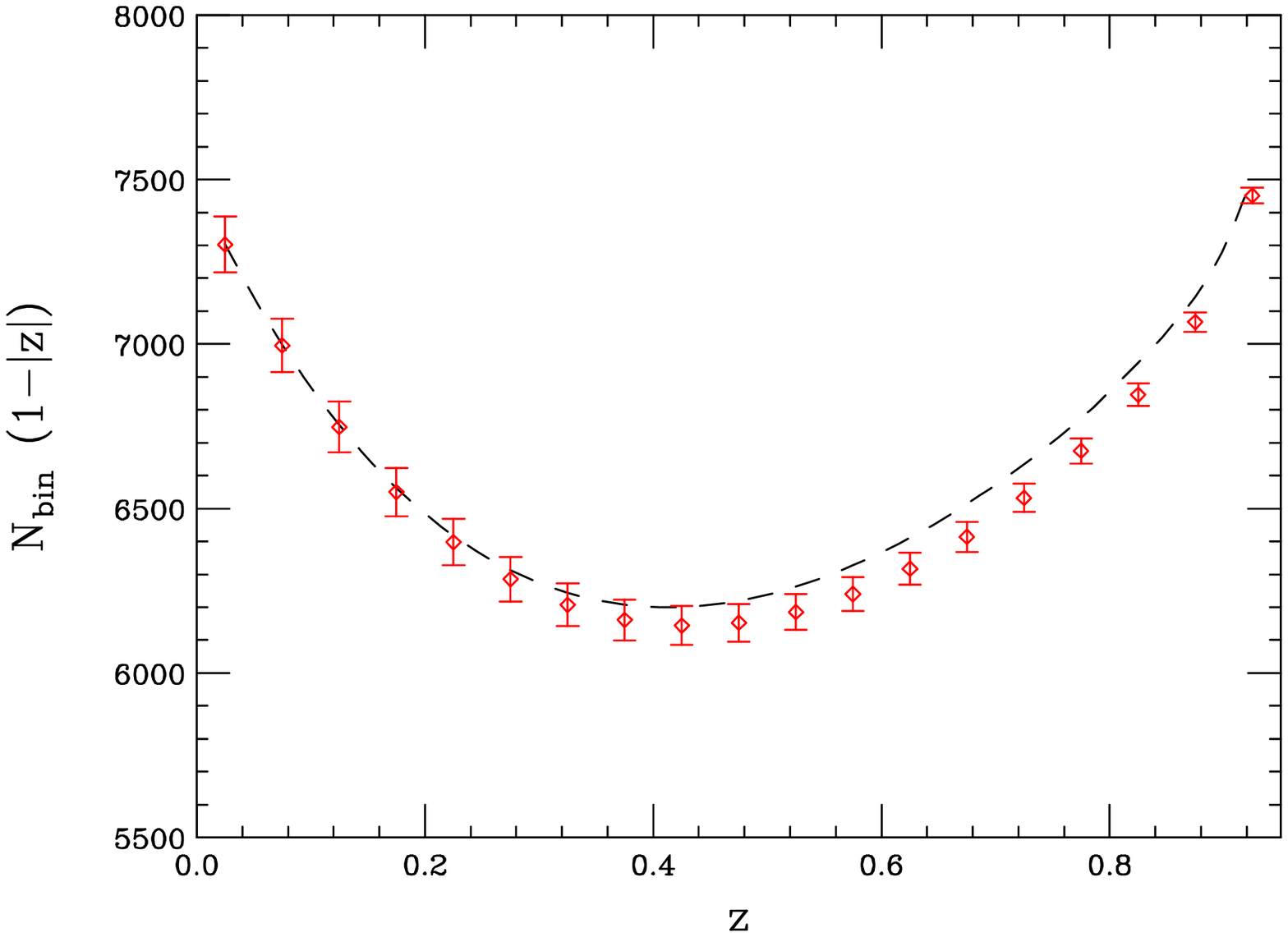,height=9cm,width=12cm,angle=0}}
\caption{$\theta$ dependence of the $e^+ e^- \rightarrow \gamma \gamma$ cross 
section for the case $\alpha = 0$.  We again use $\Lambda_{NC}={\sqrt s}= 500$ 
GeV, and a luminosity of $500 \, {\rm fb}^{-1}$.  
In the bottom panel, note that 
the number of events in each bin is scaled by $1-|z|$.}
\label{eegg0}
\end{figure}

In Figs. \ref{eeggpi2} and \ref{eegg0}  
we present the bin-integrated event rates, taking $\Lambda_{NC}=\sqrt s$
for purposes of demonstration, which show the angular 
dependences of the NC deviations for the two cases $\alpha = \pi /2$ and 
$\alpha = 0$, taking $\Lambda_{NC}={\sqrt s} =500$ GeV and a luminosity of 
$500 \, {\rm fb}^{-1}$.  For the case of $\alpha=0$ we have also scaled the
angular distribution by the factor $1-|z|$ in order to 
emphasize the deviation from the Standard Model in the peaking region.  Note
that the NC contributions lower the event rate from that expected in the SM  
in the central region.
As expected, the $\alpha=0$ case shows no $\phi$ 
dependence since the preferred direction is parallel to the beam axis, while 
the $\phi$ distribution for $\alpha=\pi/2$ exhibits a strong oscillatory
behavior.
The case $\alpha =\pi /4$, 
as well as more general choices of $\alpha$, simply extrapolates between these 
two extremes.

\begin{figure}[htbp]
\centerline{
\psfig{figure=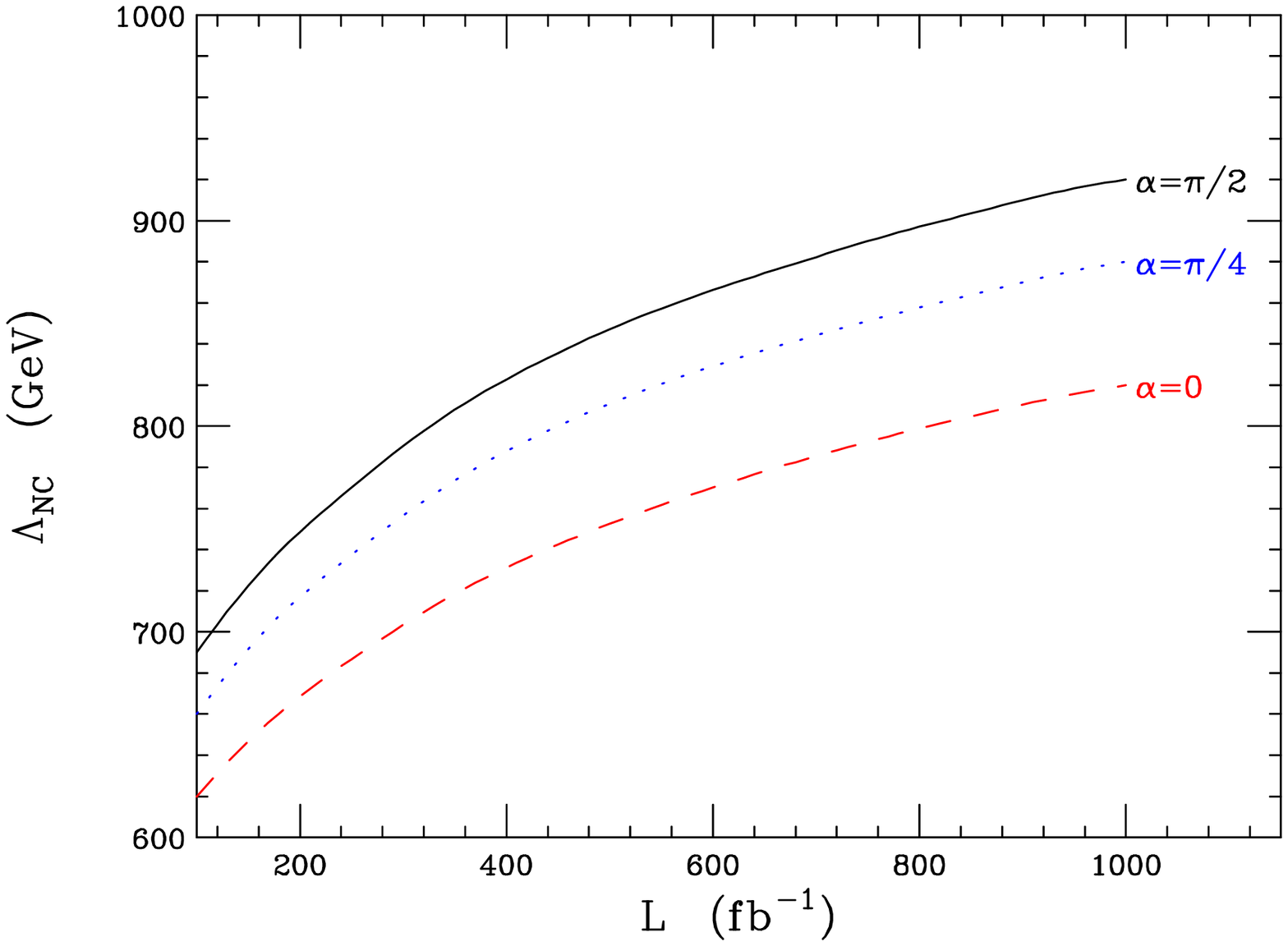,height=9cm,width=12cm,angle=0}}
\vspace*{15mm}
\centerline{
\psfig{figure=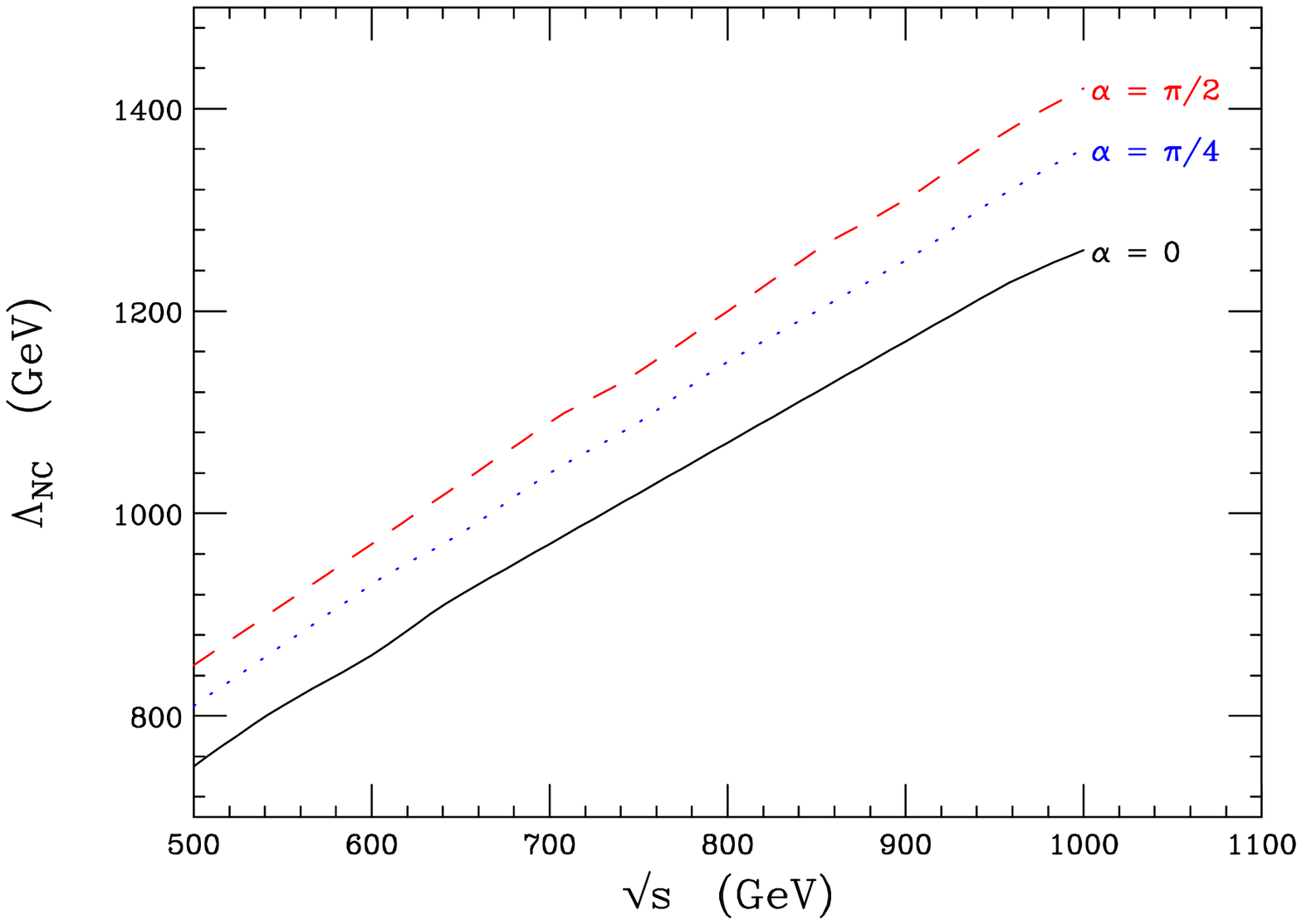,height=9cm,width=12cm,angle=0}}
\caption{$95\%$ CL bound on $\Lambda_{NC}$ from pair annihilation 
as a function of luminosity (top) and 
${\sqrt s}$ (bottom).  In the top panel we set ${\sqrt s} =500$ GeV, while 
in the bottom panel we assume a luminosity of $ 500 \, {\rm fb}^{-1}$.} 
\label{eegglim}
\end{figure}

\begin{figure}[htbp]
\centerline{
\psfig{figure=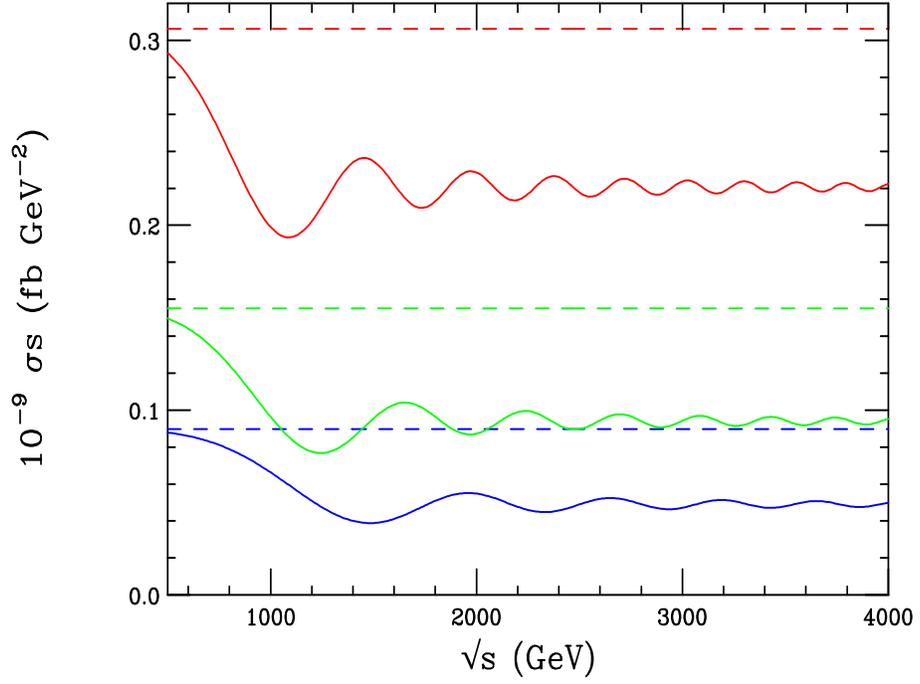,height=9cm,width=12cm,angle=90}}
\vspace*{15mm}
\centerline{
\psfig{figure=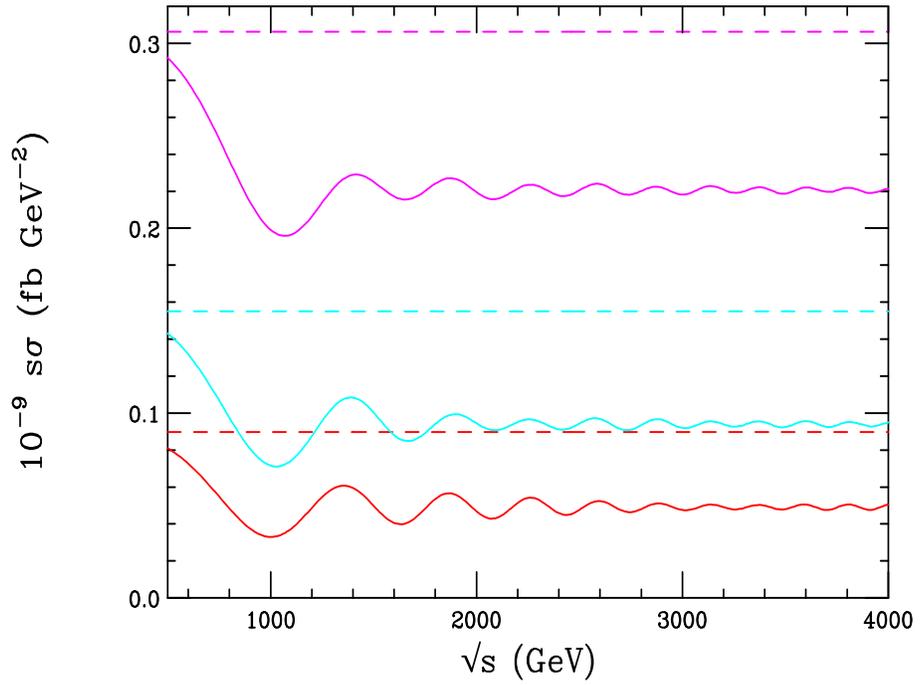,height=9cm,width=12cm,angle=90}}
\vspace*{15mm}
\caption{The scaled cross section for pair annihilation for $\alpha=0, \pi/2$
corresponding to the (top, bottom) panels, respectively.}
\label{pahe}
\end{figure}

To obtain a 95\% CL lower bound on $\Lambda_{NC}$, we perform a fit to the 
total cross section and the angular distributions employing the procedure
discussed above.  Our results 
are presented in Fig. \ref{eegglim} for three values of $\alpha$, 
where we see that the NC search reach from pair annihilation is
approximately given by $1.5\sqrt s$.  This is inferior in
comparison to that obtained in the case of Moller and Bhabha scattering,
due, in part, to the large available statistics in the latter
cases.  The scaled cross sections, after employing identical $z$ cuts as
in the previous two sections, are presented in Fig. \ref{pahe}.  Here, we see
again that the anticipated high energy behavior is realized.

\vspace*{-0.5cm}
\noindent

\noindent
\vspace*{0.1mm}

\section{$\gamma \gamma \rightarrow \gamma \gamma$ at Linear Colliders}

Future linear colliders have the option of running in a photon-photon 
collision mode\cite{GGcollider}, in which laser photons are Compton 
back-scattered off the incoming fermion beams.  The lowest 
order SM contributions arise at the 1-loop level with fermions and $W$ 
bosons propagating in the loop.  Since the exact SM calculation of this
box diagram mediated process is rather tedious\cite{GGcomp}, there exist
various approximations in the literature\cite{GGSUSY} which are
valid in the regime where the center of mass energy is large compared to
the $W$ mass.  Since this process only occurs at loop-level in the SM, it
has been proposed as a useful
test of new physics which contributes to the amplitude at the tree level in,
for example, supersymmetry\cite{GGSUSY} or quantum gravity models with large 
extra dimensions\cite{GGQG}.  In the present case, NCQED also predicts 
new contributions to 
$\gamma \gamma \rightarrow \gamma \gamma$ at tree-level, and hence we 
examine how well this process can bound $\Lambda_{NC}$.

\begin{figure}[htbp]
\centerline{
\psfig{figure=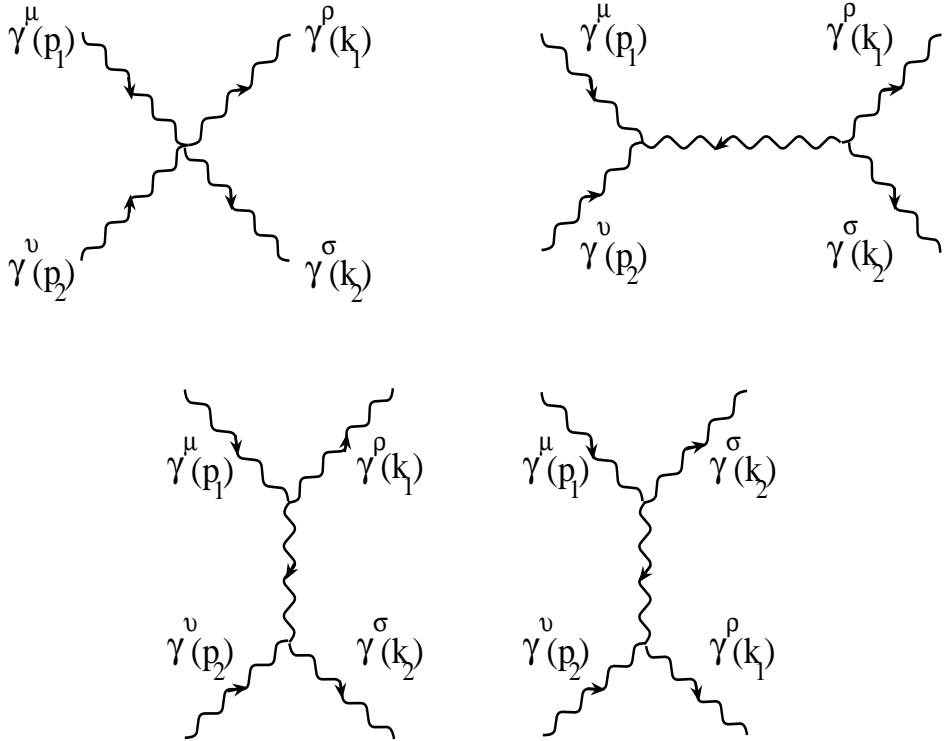,height=10.5cm,width=14cm,angle=0}}
\vspace*{0.5cm}
\caption{The tree level contributions to $\gamma \gamma \rightarrow \gamma 
\gamma$ in NCQED.}
\label{fgamma}
\end{figure}

We will consider only tree-level NC contributions since the NC generalization 
of the full electroweak SM is unknown and coupling constant suppressed.  
There are four diagrammatic contributions in this case: 
three from the $s$, $t$, and 
$u$ channels of photon exchange and one from the four-point photon coupling. 
These are presented in Fig. \ref{fgamma}.  Denoting 
the incoming photon momenta by $p_1$ and $p_2$, and the outgoing photon 
momenta by $k_1$ and $k_2$ as before, we find six non-vanishing NC 
helicity amplitudes:
\begin{eqnarray}
{\cal M}_{+--+}^{NC} &=& -32 \pi \alpha \, \frac{\hat t}{\hat s} \, 
\bigg[ \, {\rm sin}
(\frac{1}{2} p_1 \wedge k_1) \, {\rm sin}(\frac{1}{2}p_2 \wedge k_2) + 
\frac{\hat t}{\hat u} \, {\rm sin}(\frac{1}{2} p_1 \wedge k_2) \, {\rm sin}
(\frac{1}{2}p_2 \wedge k_1) \bigg] \nonumber \\
{\cal M}_{++++}^{NC} &=& 32 \pi \alpha \, \bigg[ \, 
\frac{\hat u-\hat t}{\hat s} \, 
{\rm sin}(\frac{1}{2} p_1 \wedge p_2) \, {\rm sin}(\frac{1}{2}k_1 \wedge k_2) 
\, + \, \bigg(\frac{\hat u}{\hat t} - \frac{2\hat u}{\hat s} \bigg) \, 
{\rm sin}(\frac{1}{2} p_1 
\wedge k_1) \nonumber \\ & & \times \, {\rm sin}(\frac{1}{2}p_2 \wedge k_2)   
+ \, \bigg(\frac{\hat t}{\hat u} - \frac{2\hat t}{\hat s} \bigg) \, 
{\rm sin}(\frac{1}{2} p_1 
\wedge k_2) \, {\rm sin}(\frac{1}{2}p_2 \wedge k_1) \bigg],
\end{eqnarray}
where we have made use of the relation $\hat s+\hat t+\hat u=0$ and
the $\hat s$ denotes the parton-level center-of-mass frame.  
The other four amplitudes 
are related to these by ${\cal M}_{----}^{NC}={\cal M}_{++++}^{NC}$; $\, 
{\cal M}_{+--+}^{NC}(k_1,k_2) = {\cal M}_{-++-}^{NC}(k_1,k_2) = 
{\cal M}_{+-+-}^{NC}(k_2,k_1) = {\cal M}_{-+-+}^{NC}(k_2,k_1)$.
The corresponding SM amplitudes can be found in 
Refs. \cite{GGSUSY,GGQG} and will be given in the appendix. 

\noindent

The kinematics of this process are more complicated than those of the 
previous cases.  The backscattered photons have a broad energy distribution, 
and the collision no longer occurs in the center of mass frame, \ie, the
CM and laboratory frames no longer coincide.  As NC theories violate 
Lorentz invariance, the differential cross section is no longer invariant 
under boosts along the z-axis and we are thus forced to consider this 
process in the laboratory frame.  Letting $x_1$ and $x_2$ denote the 
fraction of the fermion energy carried by each of the backscattered photons, 
the photon momenta become
\begin{eqnarray}
p_1^\mu &=& \frac{x_1 {\sqrt s}}{2}(1,1,0,0) \nonumber \\
p_2^\mu &=& \frac{x_2 {\sqrt s}}{2}(1,-1,0,0) \nonumber \\
k_1^\mu &=& E_1 (1,c_\theta,s_\theta c_\phi, s_\theta s_\phi) \nonumber \\
k_2^\mu &=& ((x_1 + x_2) \frac{{\sqrt s}}{2} -E_1,
(x_1 -x_2) \frac{{\sqrt s}}{2} -E_1 c_\theta,
-E_1 s_\theta c_\phi, - E_1 s_\theta s_\phi )\,,
\end{eqnarray}
where $E_1$ is given by 
\begin{equation}
E_1 = \frac{x_1 x_2 {\sqrt s}}{x_1 +x_2 -(x_1 -x_2) \, {\rm cos}\theta}\,.
\end{equation}
Note that the Mandelstam invariants appearing in the amplitudes are now those 
for the photon-photon center of mass frame, with, \eg, $\sqrt {\hat s} =
x_1 x_2 \sqrt s$.

We define the observable amplitudes by summing over the helicities of the 
outgoing photons:
\begin{eqnarray}
|{\cal M}_{++}|^2 &=& \sum_{ij} |{\cal M}_{++ij}|^2 \, , \nonumber \\
|{\cal M}_{+-}|^2 &=& \sum_{ij} |{\cal M}_{+-ij}|^2 \, ,
\end{eqnarray}
which also include the SM contributions.  The lab frame differential cross 
section for this process is
\begin{eqnarray}
\frac{d \sigma}{d \Omega} &=& \frac{1}{128 \pi^2 s} \, \int \, \int \, dx_1 \,
dx_2 \, \frac{E_1}{E_2} \, \frac{f(x_1) f(x_2)}{x_1 x_2} \, \bigg[ \bigg( 
\frac{1+ \xi (x_1) \xi(x_2)}{2} \bigg) \, |M_{++}|^2 \, \nonumber \\ 
& & + \, \bigg( \frac{1- \xi (x_1) \xi(x_2)}{2} \bigg) \, |M_{+-}|^2 \, 
\bigg]\,,
\end{eqnarray}
where $E_1, \, \, E_2$ denote the outgoing photon energies, $f(x)$ is the 
photon number density function, and $\xi(x)$ the helicity distribution 
function, which is 
presented in the appendix.  The distribution functions depend upon the 
variable set $ (P_{e1}, \, P_{l1}, \, P_{e2}, \, P_{l2})$, which
represent the polarizations 
of the initial fermion and laser beams.  In this paper we set $|P_e|=0.9$ 
and $|P_l|=1.0$, leaving six independent combinations: 
$(+, \, +, \, +, \, +), $ $ \, (+, \, +, \, +, \, -), $ 
$(+, \, +, \, -, \, -),$ $(+, \, -, \, +, \, -), $
$(-, \, +, \, +, \, -),$ and $(+, $ $\, -, \, -, \, -),$ where, for example,
 $(+, \, -, \, +, \, -) $ means $P_{e1}=0.9$, $P_{l1}=-1.0$, $P_{e2}=0.9$, 
and $P_{l2}=-1.0$.   We use the approximate SM amplitudes found 
in~\cite{GGSUSY,GGQG}, valid for $m_W^2 /x_p \, <1$, where $x_p$ represents 
any of the photonic Mandelstam invariants.  To validate this approximation 
we employ the cuts
\begin{equation}
|{\rm cos}(\theta)| \, \leq \, 0.8 \, , \;\;\;\; {\sqrt {0.4}} \, < \,  x_i \, 
< \, x_{max}\,.
\end{equation}
$x_{max}$ is the maximum fraction of the fermion beam energy that a 
backscattered photon can carry away; numerically, $x_{max} \approx 0.83$.  
Evaluating the wedge products in the NC amplitudes in the lab frame yields
\begin{eqnarray}
p_1 \wedge p_2 &=& \frac{-c_{01} x_1 x_2 s}{2 \Lambda^2_{NC}} \nonumber \\
p_1 \wedge k_1 &=& \frac{-x_1 E_1 {\sqrt s}}{2 \Lambda^2_{NC}} \, 
\bigg[ \, c_{01} \, (1-c_\theta ) - c_{02} s_\theta c_\phi -c_{03} s_
\theta s_\phi -c_{12} s_\theta c_\phi +c_{31} s_\theta s_\phi \, \bigg] 
\nonumber \\
p_2 \wedge k_1 &=& \frac{x_2 E_1 {\sqrt s}}{2 \Lambda^2_{NC}} \, 
\bigg[ \, c_{01} \, (1+c_\theta ) + c_{02} s_\theta c_\phi +c_{03} s_\theta 
s_\phi -c_{12} s_\theta c_\phi +c_{31} s_\theta s_\phi \, \bigg] \nonumber \\
p_1 \wedge k_2 &=& \frac{-x_1 E_1 {\sqrt s}}{2 \Lambda^2_{NC}} \, \bigg[ \, 
\frac{c_{01} x_2 {\sqrt s}}{E_1} -c_{01} (1-c_\theta ) +c_{02} s_\theta 
c_\phi + \nonumber \\ & & c_{03} s_\theta s_\phi +c_{12} s_\theta c_\phi  
-c_{31} s_\theta s_\phi \, \bigg] \nonumber \\ 
p_2 \wedge k_2 &=& \frac{-x_2 E_1 {\sqrt s}}{2 \Lambda^2_{NC}} \, \bigg[ \, 
\frac{-c_{01} x_1 {\sqrt s}}{E_1} +c_{01} (1+c_\theta ) +c_{02} s_\theta 
c_\phi +  \nonumber \\ 
& & c_{03} s_\theta s_\phi -c_{12} s_\theta c_\phi +c_{31} s_\theta s_\phi \, 
\bigg] \nonumber \\ 
k_1 \wedge k_2 &=& \frac{-E_1 {\sqrt s}}{2 \Lambda^2_{NC}} \, 
\bigg[ \, (x_1 + x_2) \{ c_{02}s_\theta c_\phi +c_{03} s_\theta s_\phi 
+c_{01} c_\theta \}  \nonumber \\ 
& & - (x_1 - x_2) \{ c_{01} - c_{12} s_\theta c_\phi +c_{31} s_\theta 
s_\phi \}  \bigg],
\end{eqnarray}
where, as before, we can interpret the $c_{\mu \nu}$ in terms of the 
directions of the background $E$ and $B$ fields, with, 
the z-axis has being defined
to be along the direction of the initial beams.  Note that in this case,
however, we have defined $p_1$ to be in the positive z-direction.  
Two important properties 
of these expressions are that the presence of both $c_{0i}$ and 
$c_{ij}$ indicates that $\gamma \gamma \rightarrow \gamma \gamma$ is 
sensitive to {\it both} space-time and space-space non-commutativity, unlike
the previously examined processes, and 
the disappearance of $c_{23}$ indicates that {\bf B} fields parallel 
to the beam axis are unobservable as in the case of Moller scattering. 
We consider three different 
possibilities: ({\it i}) $c_{01}=1$, with all others vanishing; ({\it ii}) 
$c_{03}=1$, with all others vanishing;  and
({\it iii}) $c_{12}=-1$, with all others 
vanishing.  In terms of the angular parameterization, case ({\it i}) 
corresponds to an {\bf E} field parallel to the beam axis (denoted
by $\alpha=0$ in our discussion of 
$e^+ e^- \rightarrow \gamma \gamma$), case ({\it ii}) to an {\bf E} 
field perpendicular to the beam axis ($\alpha=\pi/2$ in 
$e^+ e^- \rightarrow \gamma \gamma$), and case ({\it iii}) to a 
{\bf B} field perpendicular to the beam axis.  As noted earlier for 
$e^+ e^- \rightarrow \gamma \gamma$, $c_{02}$ and $c_{03}$ are equivalent 
up to a redefinition of $\phi$, as are $c_{12}$ and $c_{31}$.  Note, 
however, that despite their apparent similarity, the space-time and 
space-space components are {\it not} equivalent up to a redefinition of 
$\phi$.  Redefining $\phi$ in an attempt to relate $c_{03}$ and $c_{12}$  
inflicts a sign change in the amplitudes, which affects 
the interference between the SM and NC amplitudes.

\begin{figure}[htbp]
\centerline{
\psfig{figure=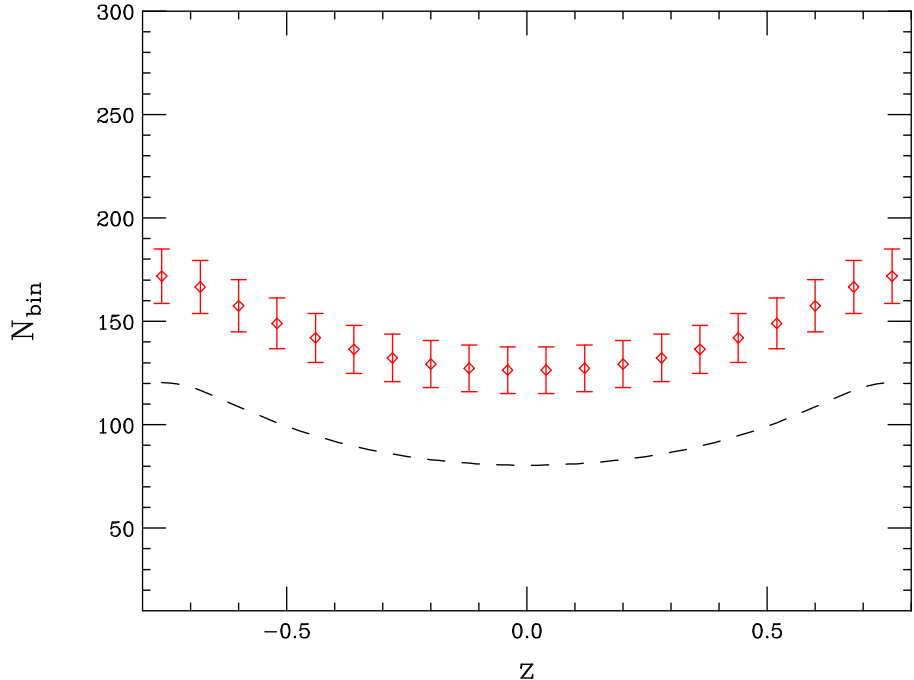,height=9cm,width=12cm,angle=0}}
\caption{Angular  dependence of the $\gamma \gamma \rightarrow \gamma \gamma$ 
cross section for the case $c_{01}=1$.  We take $\Lambda_{NC} = {\sqrt s}= 
500$ GeV, and a luminosity of $500 \, {\rm fb}^{-1}$ and employ the cuts
discussed in the text.}
\label{ggthet01}
\end{figure}

\begin{figure}[htbp]
\centerline{
\psfig{figure=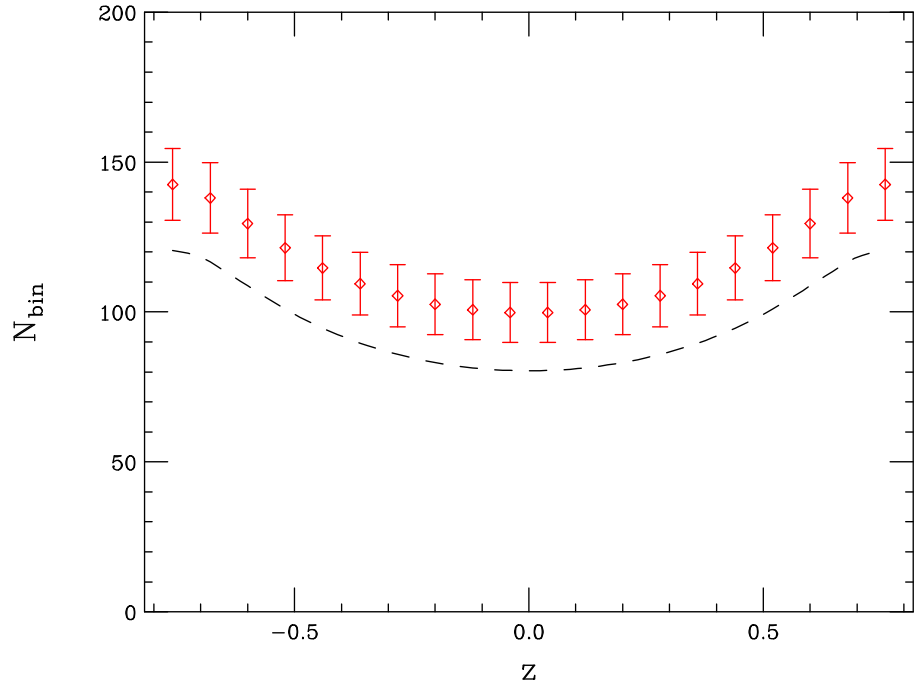,height=9cm,width=12cm,angle=0}}
\vspace*{15mm}
\centerline{
\psfig{figure=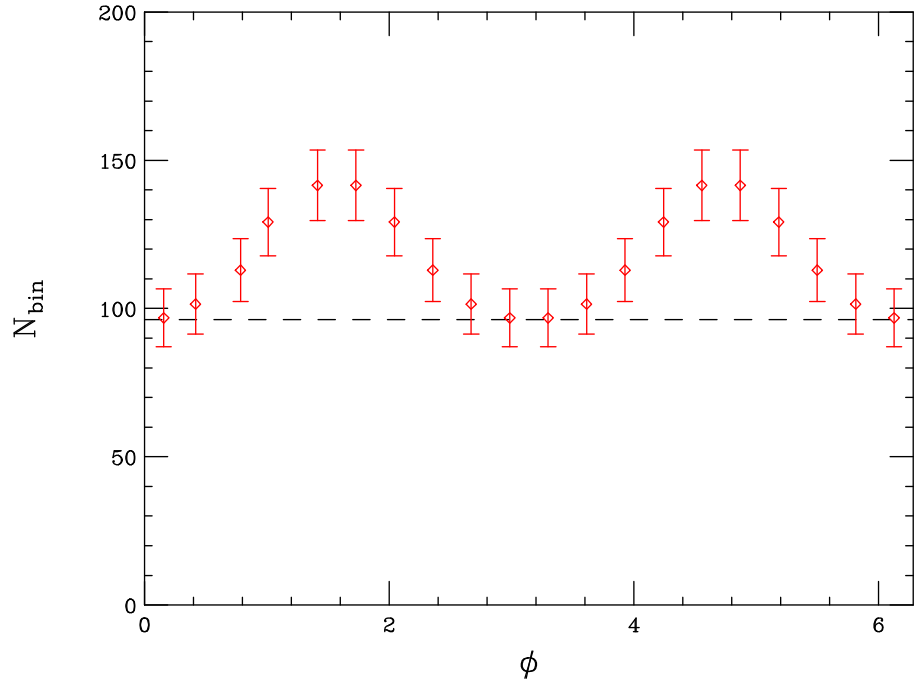,height=9cm,width=12cm,angle=0}}
\caption{$\theta$ (top) and  $\phi$ dependence (bottom) of the 
$\gamma \gamma \rightarrow \gamma \gamma$ cross section for the case 
$c_{03}=1$.  We again take $\Lambda_{NC} = {\sqrt s}= 500$ GeV, with
a luminosity of $500 \, 
{\rm fb}^{-1}$.}
\label{ggang03}
\end{figure}

\begin{figure}[htbp]
\centerline{
\psfig{figure=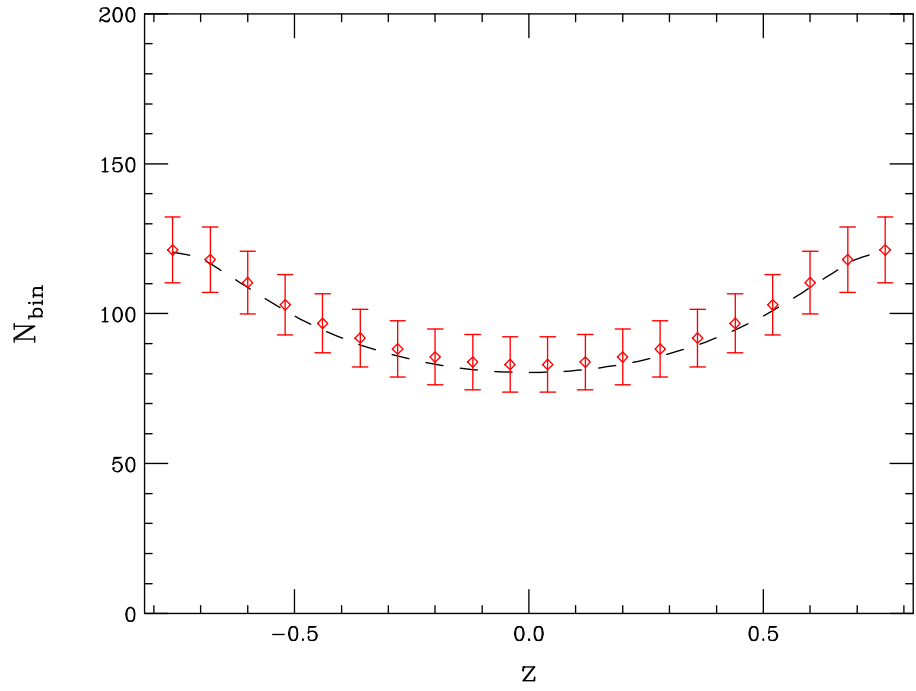,height=9cm,width=12cm,angle=0}}
\vspace*{15mm}
\centerline{
\psfig{figure=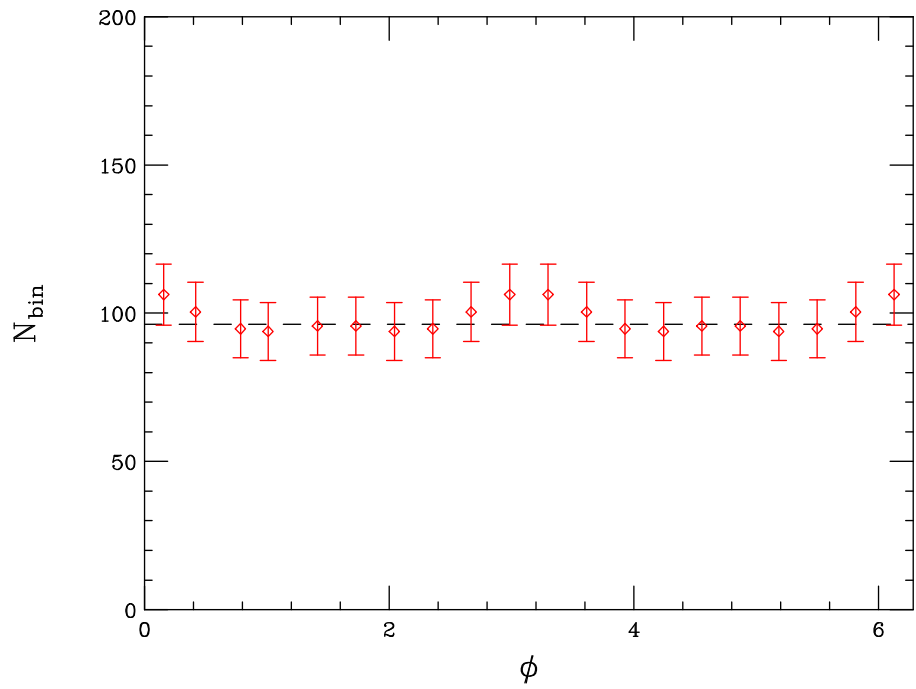,height=9cm,width=12cm,angle=0}}
\caption{Same as the previous figure, only for
$c_{12}=-1$.}
\label{ggang12}
\end{figure}

In Figs. \ref{ggthet01}, \ref{ggang03}, and \ref{ggang12}
we display the bin-integrated angular distributions 
assuming a 500 GeV \epem\ linear collider 
with an integrated luminosity of $500 \, {\rm fb}^{-1}$ and
employing the cuts discussed above.  We also take $\Lambda_{NC}=
\sqrt s$ for purposes of demonstration.
As can be seen from the figures, the effects of NC space-time yield
marked increases in both the $z$ and $\phi$ distributions over the
SM expectations, whereas this process is seen to be rather insensitive 
to space-space non-commutativity.   The NC space-space corrections also 
do not strictly increase the SM result, unlike the other two cases, due
to an interference effect between the SM and space-space NC 
contributions, and from the small magnitude 
of the NC effect in this case.

\begin{figure}[htbp]
\centerline{
\psfig{figure=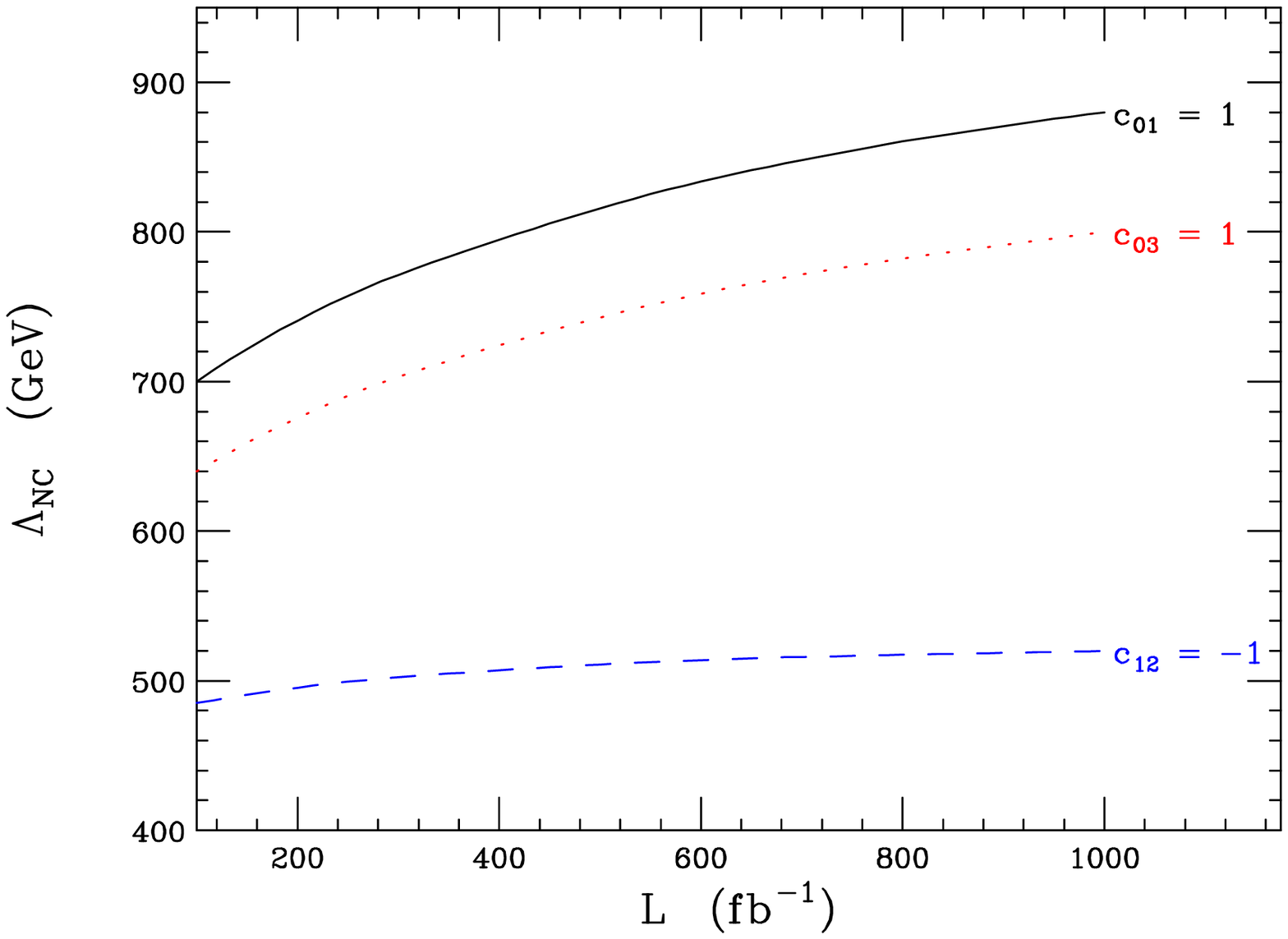,height=9cm,width=12cm,angle=0}}
\vspace*{15mm}
\centerline{
\psfig{figure=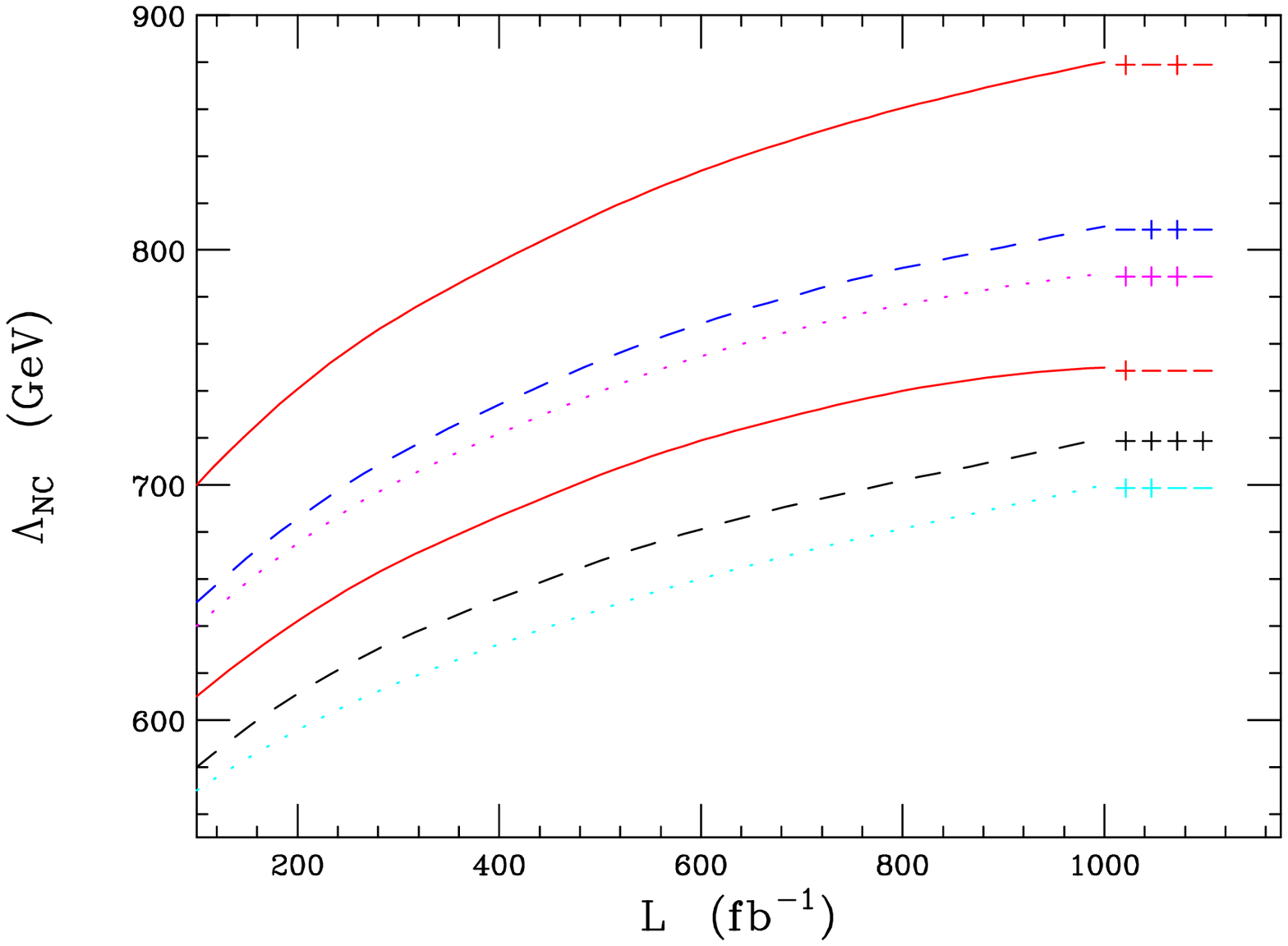,height=9cm,width=12cm,angle=0}}
\caption{$95\%$ CL bound on $\Lambda_{NC}$ from $\gamma\gamma\to
\gamma\gamma$ as a function of luminosity for ${\sqrt s} = 500$ GeV.
Top panel: the three cases of
$c_{\mu\nu}$ discussed in the text with the polarization state
$(+,-,+,-)$, and bottom panel: all polarization states with $c_{01}=1$.}
\label{ggbound}
\end{figure}

Figure \ref{ggbound} displays the $95\%$ CL search reach for
the NC scale $\Lambda_{NC}$ as a function of luminosity for
the three cases with the polarization state $(+,-,+,-)$ as well
as for the case $c_{01}$ with all polarization configurations.
As expected, $\gamma \gamma$ scattering is relatively 
insensitive to space-space non-commutativity yielding bounds that
are essentially just below  $\sqrt s$. However, in the
case of space-time NC, we see that the potential limits are 
comparable to that obtainable from pair annihilation and are of
order $1.5\sqrt s$.  2 photon scattering also nicely 
complements $e^+ e^- \rightarrow \gamma \gamma$ as one is sensitive 
to $c_{01}$ with the other depending on $c_{02}$ and $c_{03}$.

\vspace*{-0.1cm}
\noindent
\vspace*{0.1mm}

\vspace*{-0.5cm}
\noindent
\vspace*{0.1mm}

\vspace*{-0.5cm}
\noindent
\vspace*{0.1mm}

\noindent
\vspace*{0.1mm}

\section{Conclusions}

In summary, we have examined the testable nature of non-commutative quantum
field theory by analyzing various $2\to 2$ 
processes at high energy \epem\ linear
colliders.  We have parameterized the non-commutative relationship
in terms of an overall NC scale, $\Lambda_{NC}$, and an anti-symmetric
matrix $c_{\mu\nu}$ which is related to the direction of the background 
electromagnetic field present in these theories.  We have seen that
these theories give rise to modifications to QED, resulting in a non-abelian
like nature with 3- and 4-point photon self-couplings, as well as
momentum dependent phase factors appearing at each possible vertex
in NCQED.
We have seen that both Bhabha and Moller scattering are affected by
the interference momentum dependent phase factors, whereas pair annihilation 
also receives contributions from the 3-point function.  We have
also examined $\gamma\gamma\to\gamma\gamma$, which is sensitive to
both the 3- and 4-photon self-couplings.

In all the processes considered in the text, the NC affects arise at
lowest order from dimension-8 operators.  In addition, they generate
an azimuthal dependence, which is not present in the SM, due to the 
NC preferred direction in space-time.  These effects are not Lorentz
invariant, and caution must be exercised in evaluating them, both
theoretically and experimentally.

The above four processes are complementary in terms of probing the
NC parameter space.  Pair annihilation and Bhabha scattering,
together, explore the full parameter space for Space-Time 
non-commutativity, whereas Moller scattering is sensitive to 2 of
the parameters in the case of Space-Space NC.  Two photon
scattering simultaneously probes Space-Space and Space-Time NC, but
is found to be rather insensitive numerically to the Space-Space
case.  In all of these transitions, the effects of {\bf B} fields 
parallel to the beam axis are unobservable. We summarize our
results for the $95\%$ CL search reach for the NC scale in
Table \ref{summ}.  We see that NCQED can be probed to scales of
order a TeV, which is where one would expect NCQFT to become
important, if stringy effects  or if the
fundamental Planck scale are also at the TeV scale.

\begin{table}
\centering
\begin{tabular}{|c|c|c|} \hline\hline
Process & Structure Probed  & Bound on $\Lambda_{NC}$   \\ \hline
$\epem\to\gamma\gamma$   & Space-Time & $740-840$ GeV \\
Moller Scattering & Space-Space &  1700 GeV\\ 
Bhabha Scattering & Space-Time & 1050 GeV \\
$\gamma\gamma\to\gamma\gamma$ & Space-Time & $700-800$ GeV \\ 
   & Space-Space & 500 GeV \\ \hline\hline
\end{tabular}
\caption{Summary of the $95\%$ CL search limits on the NC scale
$\Lambda_{NC}$ from the various processes considered above at
a 500 GeV \epem\ linear collider with an integrated luminosity
of 500 \infb.}
\label{summ}
\end{table}

Note Added: While this paper was being completed, a related work \cite{stuff}
appeared.  There is some overlap between the processes considered in this
paper with what is contained here, however, we disagree completely with
their results.  We also note that crossing symmetry can be maintained
in NCQFT as long as one exercises care to also switch the appropriate
momenta in the wedge products.

\noindent{\Large\bf Acknowledgements}

The authors would like to thank D. Atwood, H. Davoudiasl, and
A. Kronfeld for discussions.  TGR thanks S. Gardner at the University
of Kentucky for the use of facilities.
F. P. was supported in part by a NSF Graduate Research Fellowship.

\noindent{\Large\bf Appendix}

In this appendix we present the SM amplitudes and photon distribution 
functions relevant for the process $\gamma \gamma \rightarrow \gamma \gamma$. 
For a more detailed discussion the reader is referred 
to~\cite{GGcollider,GGcomp,GGSUSY,GGQG}.

As discussed in the text,
the one loop contributions to $\gamma \gamma \rightarrow \gamma \gamma$ arise 
from $W$ boson and fermion loops.  At high energies, which we are considering, 
there is only one non-negligible independent helicity amplitude.  The 
approximate amplitudes for the $W$ contribution is
\[ 
\frac{{\cal M}_{++++}^{(W)}(s, t, u)} {\alpha^2} \approx 12 + 12 
\left(\frac{u - t}{s}\right) \left[\ln\left(\frac{-u - i\varepsilon}{m_W^2}
\right) - \ln \left(\frac{-t - i\varepsilon}{m_W^2}\right) \right]\]\[+ 16 
\left(1 - \frac{3 t u}{4 s^2}\right)\left(\left[\ln \left(\frac{-u - 
i\varepsilon}{m_W^2}\right) - \ln \left(\frac{-t - i\varepsilon}{m_W^2}
\right)\right]^2 + \pi^2 \right)\]\[+16 s^2 \left[\frac{1}{s t} \ln 
\left(\frac{-s - i\varepsilon}{m_W^2}\right) \ln \left(\frac{-t - 
i\varepsilon}{m_W^2}\right) +  \frac{1}{s u} \ln \left(\frac{-s - 
i\varepsilon}{m_W^2}\right) \ln \left(\frac{-u - i\varepsilon}{m_W^2}\right) 
\right.\]
\begin{equation}
+ \left.\frac{1}{t u} \ln \left(\frac{-t - i\varepsilon}{m_W^2}\right) 
\ln \left(\frac{-u - i\varepsilon}{m_W^2}\right)  \right],
\label{Wamp}
\end{equation}
where $\alpha \approx 1/137$, $m_W$ represents the mass of the $W$ boson 
and $\varepsilon$ is a small positive quantity defining the branch 
cut prescription.  The fermion contribution gives rise to the approximate
amplitude
\[ \frac{{\cal M}_{++++}^{(f)}(s, t, u)}{\alpha^2 Q_f^4} \approx -8 - 8 
\left(\frac{u - t}{s}\right) \left[\ln \left(\frac{-u - i\varepsilon}{m_f^2}
\right) - \ln \left(\frac{-t - i\varepsilon}{m_f^2}\right) \right]\]
\begin{equation}
 - 4 \left(\frac{t^2 + u^2}{s^2}\right)\left(\left[\ln \left(\frac{-u - 
i\varepsilon}{m_f^2}\right) - \ln \left(\frac{-t - i\varepsilon}{m_f^2}
\right)\right]^2 + \pi^2 \right),
\label{famp}
\end{equation}
where $Q_f$ is the fermion charge in units of the positron charge, and 
$m_f$ is the mass of the fermion in the loop.  The other amplitudes 
 are related to these by
\begin{equation}
{\cal M}_{+-+-}(s,t,u) \, =\,  {\cal M}_{+--+}(s,u,t) \, = \, 
{\cal M}_{++++}(u,t,s).
\end{equation}

We now present the expressions for the photon distributions.
We define the auxiliary functions
\begin{equation}
C(x) \equiv \frac{1}{1 - x} + (1 - x) - 4 r (1 - r) - P_e \, P_l \, r \, z 
(2 r - 1) (2 - x),
\label{C(x)}
\end{equation}
where $r = x/[z(1 - x)]$, and
\[
\sigma_{_{C}} = \left(\frac{2 \pi \alpha^2}{m_e^2 z}\right) \left[\left(1 - 
\frac{4}{z} -\frac{8}{z^2}\right) \ln (z + 1) + \frac{1}{2} + \frac{8}{z} - 
\frac{1}{2 (z + 1)^2}\right]\]
\begin{equation}
+ P_e \, P_l \left(\frac{2 \pi \alpha^2}{m_e^2 z}\right) \left[\left(1 + 
\frac{2}{z}\right) \ln (z + 1) - \frac{5}{2} + \frac{1}{z + 1} - 
\frac{1}{2 (z + 1)^2}\right].  
\end{equation}
Here $z$ is a variable describing the laser photon energy; varying $z$ 
affects the value of 
$x_{max}$, the maximum value of the fermion beam energy that the backscattered 
photons can acquire.  We set $z =2 (1+{\sqrt 2})$ in our analysis, which 
maximizes $x_{max}$.  In terms of these functions the photon number and 
helicity distribution functions take the form
\begin{eqnarray}
f(x, P_e, P_l; z) &=& \left(\frac{2 \pi \alpha^2}{m_e^2 z 
\sigma_{_{C}}}\right) C(x)  \\ 
\xi(x, P_e, P_l; z) &=& \frac{1}{C(x)}\left\{P_e \, \left[\frac{x}{1 - x} 
+ x (2 r - 1)^2\right]
 - P_l \, (2r - 1)\left(1 - x + \frac{1}{1 - x}\right)\right\}\,. \nonumber
\end{eqnarray}

\newpage

%
\def\MPL #1 #2 #3 {Mod. Phys. Lett. {\bf#1},\ #2 (#3)}
\def\NPB #1 #2 #3 {Nucl. Phys. {\bf#1},\ #2 (#3)}
\def\PLB #1 #2 #3 {Phys. Lett. {\bf#1},\ #2 (#3)}
\def\PR #1 #2 #3 {Phys. Rep. {\bf#1},\ #2 (#3)}
\def\PRD #1 #2 #3 {Phys. Rev. {\bf#1},\ #2 (#3)}
\def\PRL #1 #2 #3 {Phys. Rev. Lett. {\bf#1},\ #2 (#3)}
\def\RMP #1 #2 #3 {Rev. Mod. Phys. {\bf#1},\ #2 (#3)}
\def\NIM #1 #2 #3 {Nuc. Inst. Meth. {\bf#1},\ #2 (#3)}
\def\ZPC #1 #2 #3 {Z. Phys. {\bf#1},\ #2 (#3)}
\def\EJPC #1 #2 #3 {E. Phys. J. {\bf#1},\ #2 (#3)}
\def\IJMP #1 #2 #3 {Int. J. Mod. Phys. {\bf#1},\ #2 (#3)}
\def\JHEP #1 #2 #3 {J. High Energy Phys. {\bf#1},\ #2 (#3)}

\end{document}